\documentclass{article}
\usepackage{amsfonts,amsbsy}
\usepackage{mathtools}
\usepackage{setspace}
\usepackage[framemethod=tikz]{mdframed} 
\usepackage{caption}
\usepackage{float}
\usepackage{enumitem}
\usepackage{booktabs}
\usepackage{graphicx}
\usepackage{longtable}
\usepackage{bm}
\usepackage{color,amsthm,physics}
\usepackage{comment}
\usepackage{hyperref} 
\usepackage{longtable}
\usepackage{booktabs}
\usepackage{setspace}

\usepackage[utf8]{inputenc}
\usepackage[english]{babel}

\usepackage{rotating}
\usepackage{caption}
\usepackage{subcaption}

\theoremstyle{remark} 
\newmdtheoremenv[linewidth=0,nobreak=true]{theorem}{Theorem}
\newmdtheoremenv[linewidth=0,nobreak=true]{intuition}{Intuition}
\newmdtheoremenv[linewidth=0,nobreak=true]{definition}{Definition}
\newmdtheoremenv[linewidth=0,nobreak=true]{remark}{Remark} 
\newmdtheoremenv[linewidth=0,nobreak=true]{lemma}{Lemma} 
\newmdtheoremenv[linewidth=0,nobreak=true]{algorithm}{Algorithm} 
\newmdtheoremenv[linewidth=0,nobreak=true]{corollary}{Corollary}

\newcount\Comments  
\usepackage{color}
\definecolor{darkgreen}{rgb}{0,0.5,0}
\definecolor{purple}{rgb}{1,0,1}
\newcommand{\kibitz}[2]{\ifnum\Comments=1\textcolor{#1}{#2}\fi}

\newif\ifworkingversion

\theoremstyle{definition}

\theoremstyle{plain}

\usepackage[noabbrev,nameinlink,capitalise]{cleveref}
\crefformat{equation}{#2#1#3}
\crefrangeformat{equation}{(#3#1#4)--(#5#2#6)}
\AtBeginDocument{\renewcommand{\ref}[1]{\cref{#1}}}
\usepackage[section]{placeins}

\usepackage[sectionbib,sort]{natbib}
\usepackage[section]{placeins}
\usepackage{dirtytalk}

\newcommand*{\SuperScriptSameStyle}[1]{%
  \ensuremath{%
    \mathchoice
      {{}^{\displaystyle #1}}%
      {{}^{\textstyle #1}}%
      {{}^{\scriptstyle #1}}%
      {{}^{\scriptscriptstyle #1}}%
  }%
}

\newcommand*{\oneS}{\SuperScriptSameStyle{*}}
\newcommand*{\twoS}{\SuperScriptSameStyle{**}}
\newcommand*{\threeS}{\SuperScriptSameStyle{*{*}*}}

\newrobustcmd{\B}{\bfseries}

\usepackage{multirow}
\usepackage{hyperref}
\usepackage{array}

\title{A Comparison of Cryptocurrency Volatility-benchmarking New and Mature Asset Classes}

\author{Alessio Brini\footnote{Corresponding author} \quad \emph{alessio.brini@duke.edu}, \and Jimmie Lenz \quad \emph{jimmie.lenz@duke.edu}}

\date{%
  Digital Asset Research and Engineering Collaborative (DAREC) Lab, Pratt School of Engineering, Duke University \\ 
  305 Teer Engineering Building Box 90271, Durham, NC 27708, United States%
}

\begin{document}

\maketitle
\onehalfspacing


\section*{Abstract}
The paper analyzes the cryptocurrency ecosystem at both the aggregate and individual levels to understand the factors that impact future volatility. The study uses high-frequency panel data from 2020 to 2022 to examine the relationship between several market volatility drivers, such as daily leverage, signed volatility and jumps. Several known autoregressive model specifications are estimated over different market regimes, and results are compared to equity data as a reference benchmark of a more mature asset class. The panel estimations show that the positive market returns at the high-frequency level increase price volatility, contrary to what is expected from the classical financial literature. We attributed this effect to the price dynamics over the last year of the dataset (2022) by repeating the estimation on different time spans. Moreover, the positive signed volatility and negative daily leverage positively impact the cryptocurrencies' future volatility, unlike what emerges from the same study on a cross-section of stocks. This result signals a structural difference in a nascent cryptocurrency market that has to mature yet. Further individual-level analysis confirms the findings of the panel analysis and highlights that these effects are statistically significant and commonly shared among many components in the selected universe.

\section{Introduction}\label{Sec:intro}

Cryptocurrencies have continuously gained popularity since Bitcoin (BTC), the first and most well-known cryptocurrency, was created in 2009, right after the 2008 financial crisis, as a response to traditional financial institutions' perceived lack of trust. Since its advent, many other cryptocurrencies have emerged, with around a 400\% growth in the number of traded coins and tokens on multiple exchanges since the beginning of 2020\footnote{https://www.statista.com/statistics/863917/number-crypto-coins-tokens/}. The rise of digital currencies can be attributed to several factors due to the heterogeneous nature and use cases each has. One reason behind this sharp increase in interest is their potential for enhancing financial freedom and removing intermediaries from financial transactions. Individuals can make peer-to-peer transactions with cryptocurrencies without a central authority or intermediaries. This characteristic has made them particularly appealing to those concerned about government control and censorship in traditional financial systems. Another relevant factor contributing to the popularity of cryptocurrencies is their growing acceptance and use as a form of payment by merchants and businesses in less-developed countries where only some have access to a traditional bank account. Furthermore, the development of decentralized finance (DeFi) platforms, such as decentralized exchanges or lending protocols, has provided additional motivation to enter the cryptocurrency ecosystem for those who may have previously been excluded from traditional financial systems. However, beyond those premises, the cryptocurrency market is still relatively immature and subjected to substantial uncertainty and sudden failures. In that regard, the cryptocurrency market historically showed high volatility traits \citep{bouri2019co,bouri2020volatility,bouri2020bitcoin} and potential for significant returns, resulting in being attractive for retail investors. Therefore, despite its growing popularity, the cryptocurrency market, if viewed in the context of price versus more mature asset classes, is notoriously unstable, with frequent and substantial fluctuations in value. 

A known stylized fact of traditional financial market is the asymmetric impact of financial returns on the future volatility of the assets, referred to as the leverage effect \citep{black1976studies}, in which negative returns increase the asset's volatility more than positive returns \citep{bollerslev2006leverage,bollerslev2006volatility,carr2017leverage}. The presence of the leverage effect in a financial return series is often motivated by the market participants' response to adverse shocks. The increased volatility due to negative returns can start a self-reinforcing mechanism that leads to sudden asset \say{firesales}, exacerbating the uncertainty even further. Given the novelty of cryptocurrencies as an asset class, it is essential to determine if the asymmetric effect of volatility, commonly seen in mature markets, is also present here. The unclear classification of cryptocurrencies on whether they align with established asset classes or stand apart \cite{baur2018bitcoin,corbet2018exploring}, compounded by the evolving regulatory landscape \cite{shanaev2020taming}, makes this investigation particularly crucial to provide a starting point for a sound and data-driven comparison that can reflect how the market perceives cryptocurrencies. Some previous work tested the asymmetric effect of cryptocurrency returns at daily frequency \citep{phillip2018new,baur2018asymmetric,brini2022assessing} on a more or less comprehensive cross-section of cryptocurrencies. The results at the daily level detect the absence of a statistically significant leverage effect in such a market, showing an inversion of the asymmetric effect of returns. In this case, investors exploit market crashes as a buying opportunity to enter at a perceived discount. This absence of panicking behaviors among market participants indifferent to the sign of the returns in terms of reaction size let them enter the market without understanding the asset class, for the most part, investing money because this is considered a new investment with many potential opportunities.

This paper extends the analysis of the price volatility inherent in the cryptocurrency market at a higher frequency level, exploring its dynamics and explaining its main drivers. The developing interest in cryptocurrencies from regulatory bodies partially stems from the lack of clarity regarding their classification as an asset class \citep{corbet2019cryptocurrencies}. A detailed study of market behaviors, such as the asymmetric effects on volatility, can contribute to a deeper understanding of the fundamental characteristics of cryptocurrencies. By comparing these behaviors with those observed in established asset classes, we can provide insights that may assist in developing appropriate regulatory measures other than better-informed investment choices. Such insights are particularly valuable given the cryptocurrency markets' rapid growth and the increasing number of market participants. Therefore, the outcomes of such analysis can inform the direction of regulatory policies, ensuring they are based on empirical evidence of the asset class’s inherent nature. Our analysis aims to understand the cryptocurrency's volatility dynamics that emerge from high-frequency data due to different market participants since retail investors usually operate on daily if not lower frequencies \citep{auer2022crypto}. Previous works \citep{jha2020regime,ftiti2021cryptocurrency} have already studied the volatility dynamics of cryptocurrency at a frequency higher than daily. However, these analyses are restricted to a single or few cryptocurrencies, and the insights are detected and described only within the realm of the crypto ecosystem. We extend the analysis in three directions to further elaborate on these aspects. First, we consider historical high-frequency cryptocurrency data represented by a large cross-section of coins and tokens to understand the volatility dynamics of the global cryptocurrency market. We significantly extend the analysis of cryptocurrency volatility by enlarging the analyzed universe, which includes 87 cryptocurrencies. Secondly, we provide a basis for comparison by performing the same volatility analysis on a cross-section of stocks, representing a proxy of a more established and mature asset class. The selected stocks come from the NASDAQ technology index (NDXT), whose returns are positively correlated with the return of cryptocurrencies in the past years \citep{goodell2021diversifying}. By focusing the volatility analysis on multiple and larger cross-sections, we can fulfill a gap of interpretability available in the literature that often overlooks a thoughtful comparison with traditional financial markets, even knowing that cryptocurrencies often exhibit a certain level of correlation with them. Lastly, the period covered by our dataset, from 2020 to 2022, allows us to appreciate the significance of our selected volatility drives under the latest bullish and bearish market regimes that characterize the cryptocurrency market. Such a period has been crucial for developing the cryptocurrency industry, which welcomed more institutional investors among the active participants and experienced astounding growth followed by sudden drawdowns. Previous works focus on the analysis of periods pre-2020 when the market dynamics and cryptocurrency adoption were less developed than in the subsequent years. We consider combining these three elements essential to disentangle the peculiar and common traits of cryptocurrencies as a whole with respect to more traditional assets.

To this end, we adopt the framework of \citep{patton2015good} by obtaining volatility estimators from minute-level price data and fitting a set of renowned interpretable econometric models to describe the volatility patterns over time. Estimating volatility from historical data is crucial in financial econometrics, as it helps assess risk and make informed investment decisions. Various methods have been developed to estimate volatility, including historical, implied, and realized volatility. Historical volatility is often indicated as a measure of the standard deviation of an asset's returns over a specified period, usually calculated daily, while implied volatility is derived from the prices of options on the underlying asset. The latter is a forward-looking measure incorporating market expectations about future volatility by assuming market efficiency and exact price discovery for the options. In our work, we adopt the realized volatility as an estimator for historical volatility, which employs intraday data. Unlike the volatility calculated daily, realized volatility allows us to capture the past volatility dynamics more accurately and can account for sudden changes in market conditions. Being calculated on a larger amount of data points, it is also less sensitive to outliers and price gaps, and it is also not affected by assumptions on the market as the implied volatility. The choice is further motivated by the possibility of accounting for the effects of market microstructure noise, which can distort the volatility estimation. This is particularly important in highly volatile markets such as cryptocurrency, where small deviations from the true volatility can significantly affect investment decisions.

Studying the volatility dynamics of the cryptocurrency market is important for several scopes, among which managing risk effectively and hedging losses \citep{almeida2022systematic}, enhancing market efficiency through trading \citep{fang2022cryptocurrency} and designing derivatives \citep{soylemez2019cryptocurrency,akyildirim2020development,soska2021towards}. In particular, the presence of derivatives in the market modulates the volatility transmissions among centralized cryptocurrency exchanges \citep{alexander2022role,badenhorst2018effect,beneki2019investigating}.  The cryptocurrency market is relatively new and unregulated, meaning price discovery can be highly volatile and erratic. Under this view, it is a known fact that the cryptocurrency market still contains a lot of inefficiencies compared to more established financial markets (see \cite{al2020cryptocurrency} for an analysis of how a high volatility period harms the market efficiency for cryptocurrencies). Studying the volatility dynamics of cryptocurrencies can help traders identify such inefficiencies and improve the amount of information reflected by the prices when they take advantage of them. Hence, it makes the study of volatility patterns crucial for market stability. Moreover, the cryptocurrency market's relatively young age, the lack of regulation, and the high level of speculation make a proper understanding of the volatility dynamics a pressing problem for the industry aiming to gain market stability.

The rest of the paper is organized so that \ref{Sec:lit} illustrates the related literature and \ref{Sec:methdata} introduces the methodologies to estimate the volatility, provides details regarding the structure of the analyzed data, and explains the employed autoregressive models in detail. \ref{Sec:result} includes the results of the empirical analysis, from the model estimation using panel data to the robustness check by repeatedly fitting the same model specification over a different time window. In addition, we carry out the individual estimation of the same model for every entity in the cryptocurrency cross-section to disentangle the aggregate results and validate them. Then \ref{Sec:discussion} discusses the key findings and the implications of our results for the cryptocurrency ecosystem and \ref{Sec:conclusion} provides the takeaways from the volatility analysis.

\section{Related Literature}\label{Sec:lit}
The literature relevant to this study bifurcates into two distinct yet interconnected domains. The first is the foundational body of work on volatility modeling in finance, which has been a key topic in financial econometrics. This part of the literature particularly explores modeling the volatility of equity returns through variants of the autoregressive conditional heteroskedasticity (ARCH) model \citep{engle1987estimating} and its generalized (GARCH) version \citep{bollerslev1987conditionally} (see also \cite{bollerslev1994arch,andersen2006volatility} for a comprehensive overview). The use of high-frequency data substantiates the discoveries regarding volatility modeling, among which the impact of signed returns on future volatility, referred to as the leverage effect, represents a well-established empirical regularity in financial market data. \cite{bollerslev2006leverage,barndorff2008measuring,chen2011news} highlights the effect of negative equity returns on increasing future volatility.

The second strand of literature explores how these studies analyze equity volatility patterns by computing high-frequency volatility estimators such as the realized variance \citep{andersen2001distribution} and the bipower variation \cite{barndorff2006econometrics}. Alternative volatility estimators like realized semivariance \citep{barndorff2008measuring} dissect the realized variance measure, isolating the elements attributable solely to positive and negative high-frequency returns. The significance of these estimators lies in their ability to retain the critical information encoded in the sign of returns, which is otherwise lost when volatility is measured through the sum of squared or absolute value returns.

The econometrics literature has shown how high-frequency price data can improve the estimation and predictability of the volatility for a cross-section of equities \citep{patton2015good,bollerslev2020good}. These studies leverage the heterogeneous autoregressive (HAR) model proposed by \cite{corsi2009simple} and elaborate on its variants to account for the direction of the volatility movements. All the models use realized variance estimators and their variants obtained from high-frequency data.

The choice of the realized variance as an estimator is motivated by efficiency, being an estimator with low variance among all the unbiased ones, and by consistency, approaching the true value of the integrated variance process when the number of observations increases. In addition, it is easy to compute, as it just requires price data sampled at an intraday frequency. It is also flexible because it can be generalized to other volatility measures that reflect specific aspects of the volatility process, such as the realized semivariance to assess the downside risk of an asset (see \cite{ait2014high} for more details). Other volatility estimators do not consider the market microstructure dynamics (daily historical variance) or are too complex and not easily adaptable to high-frequency data (GARCH-like models). Those choices, such as the Parkinson estimator \citep{parkinson1980extreme} and its extension \cite{garman1980estimation}, systematically underestimate the magnitude of the volatility, calculating it on daily data and assuming the price process as a continuous process without jump components.

The analysis of volatility in the cryptocurrency market has attracted significant attention from researchers and practitioners due to the highly volatile nature of this market. The use of those volatility estimators is relatively new to the field of cryptocurrencies, where the realized variance estimator is calculated and modeled on just BTC \citep{hu2019risk, yu2019forecasting,shen2020forecasting}. Common findings are the presence of an inverse leverage effect that impacts the estimation of future volatility and the role of jumps in shaping the Bitcoin volatility dynamics \citep{chaim2018volatility,charles2019volatility}. Previous studies have explored various aspects of cryptocurrency market volatility daily, including its determinants, spillover effects, and forecasting accuracy. For instance, \cite{baur2018asymmetric} examined the volatility of several cryptocurrencies and found that different factors, such as market capitalization and trading volume, drive their volatility. \cite{zhang2018some} explores the stylized facts of the daily returns of eight cryptocurrencies, finding heavy-tailed distributions. In contrast, \cite{phillip2018new} study the stylized facts of a larger cross-section of cryptocurrency daily returns. In particular, the literature analyzed the time-varying relationship between financial returns and their volatility at the daily level is usually investigated through a GARCH framework. Many works \citep{liu2019volatility,fakhfekh2020volatility,wajdi2020asymmetric} retrieve an inverted leverage effect in the cryptocurrency market using several variants of the original GARCH model on daily data. \cite{chu2017garch} apply GARCH-like models to fit the best volatility model for prediction, while \cite{mostafa2021gjr} use asymmetric GARCH models for risk management assessment. \cite{alqaralleh2020modelling} compares the predictive power of nonlinear GARCH-type models for a small subset of cryptocurrency assets similar to \cite{lopez2022cryptocurrency}, which retrieves an inverse leverage effect by examining the return distribution. \cite{chaim2019nonlinear} employ a multivariate stochastic volatility model with discontinuous jumps to mean returns and volatility and signals larger transitory mean jumps starting from 2017 as an effect of shifts in cryptocurrencies return dynamics. \cite{dyhrberg2016bitcoin} identifies Bitcoin as useful in risk management and ideal for risk-averse investors in anticipation of negative shocks to the market through an asymmetric GARCH model. \cite{gradojevic2021volatility} analyze the volatility cascades for some highly capitalized cryptocurrencies. \cite{ji2019dynamic} explores the volatility connectedness, retrieving an asymmetric effect depending on the direction of the returns.

Specifically, \cite{jha2020regime,ftiti2021cryptocurrency} explores the analysis of cryptocurrency volatility by using intraday data of a small number of coins and tokens. They discovered the inverse leverage effect at high frequency based on data before 2020. Similarly, \cite{naeem2022good,yousaf2020discovering} study spillover effect on the volatility using generalized vector autoregressive model versions. \cite{mensi2021high} focuses on the portfolio management implication of high-frequency volatility patterns. \cite{katsiampa2019high} explores the volatility co-movements of eight cryptocurrencies on hourly data, while \cite{sensoy2021high,ampountolas2022cryptocurrencies} explore the spillover effects on the volatility of high-frequency cryptocurrency returns. \cite{ji2021realised,zhang2019stylised} investigate the stylized fact of high-frequency return in terms of the Hurst Exponent. \cite{katsiampa2019empirical} find asymmetric effects between good and bad news among cryptocurrencies.

\cite{baur2018asymmetric,brini2022assessing} analyze the inversion of leverage effect for the cryptocurrency market at the daily level by linking the increased excitement and speculative behavior when the market is rising with behavioral traits of the average uninformed investors, such as the fear of missing out (FoMO) \citep{mcginnis2004social}. \cite{wang2023fomo} explore the FoMo idea specifically on BTC finding positive asymmetric volatility behavior in the Bitcoin market. \cite{huang2022leverage} detail the contribution of jumps in cryptocurrency price series to the inverse leverage effect. \cite{panagiotidis2022volatility} retrieves the inversion of the leverage effect for a wide cross-section of cryptocurrencies using a GARCH-like framework. \cite{ardia2019regime} find evidence of regime changes in the GARCH volatility process for BTC, highlighting the temporary and transient nature of such volatility dynamics. \cite{katsiampa2019volatility} retrieve volatility spillover effect for various cryptocurrencies using a BEKK-MGARCH model.

\section{Methodology and Data}\label{Sec:methdata}
This Section outlines the steps needed to perform the empirical analysis in the following Section. We first describe the methodology for obtaining the daily volatility estimators from high-frequency data. Then, we present the collected data outlining descriptive statistics of the universe of the two asset classes, cryptocurrency, and equity. The final part of the Section describes the model specifications employed and the estimation techniques to gain insights into the drivers of future volatility.
\subsection{Volatility Estimator} \label{Subsec:estimators}
All the estimated econometric models in our empirical analysis are based on high-frequency volatility estimators, making it essential to start by describing the underlying assumptions and the choices to compute such estimators. As outlined in \ref{Sec:intro}, volatility estimators based on high-frequency data are more expressive of the true variation of the cryptocurrency returns and are less sensitive to the presence of outliers, making it a suitable choice to study a market subjected to sudden and irrational movements. Hence, we approach modeling the future volatility by first describing the employed volatility estimators. Let $p_t$ a continuous-time stochastic process of log prices
\begin{equation}\label{Eq:logp}
p_t=\int_0^t \mu_s \mathrm{~d} s+\int_0^t \sigma_s \mathrm{~d} W_s+J_t
\end{equation}
including two continuous components and a jump component. We call $\mu$ a locally bounded drift process, $\sigma$ a strictly positive cadlag process, and $J$ a jump process. The quadratic variation of the process in Eq. \ref{Eq:logp} is given by 
\begin{equation}\label{Eq:qv}
[p, p]=\int_0^t \sigma_s^2 \mathrm{~d} s+\sum_{0<s \leq t}\left(\Delta p_s\right)^2
\end{equation}
where $\Delta p_s=p_s-p_{s-}$ captures the jump component.

The realized variance (RV) is an estimator of Eq. \ref{Eq:qv} such that
\begin{equation}\label{Eq:RV}
R V_t=\sum_{i=1}^n r_{t, i}^2 \stackrel{p}{\longrightarrow}[p, p] \text {, as } n \longrightarrow \infty
\end{equation}
where $r_{t, i}=p_t-p_{t-1}$ is the return calculated as the log price difference. The estimator has been shown to converge in probability to the quadratic variation as the time interval between each equally spaced observation becomes small \citep{andersen2003modeling}. The RV represents the quadratic variation of a process as the sum of squared returns sampled at a high frequency. For instance, the $n$-sample RV measure is obtained from $n+1$ equally spaced sample of the price process. In general, using higher frequency data to calculate realized variance can result in more accurate estimates of volatility, as it captures more of the price fluctuations that occur within the time period being analyzed. However, there are trade-offs to consider when using higher-frequency data since a large number of data points can lead to more noise in the data and make it more difficult to identify the true volatility of the asset. Hence, the optimal sampling frequency to compute realized variance is a trade-off between bias and efficiency \citep{mcaleer2008realized}. That optimal has been identified as the 5-minute frequency to provide accuracy in estimation and avoid microstructure noises such as bid-ask bounce and price discreteness \citep{andersen1998answering}.

Another powerful estimator of the variance process is the bipower variation (BV), which, unlike the RV, converges in the limit to just the continuous component of the quadratic variation, called integrated variance (IV)
\begin{equation}\label{Eq:BV}
    BV=\frac{\pi}{2} \sum_{i=2}^n\left|r_i\right|\left|r_{i-1}\right| \stackrel{p}{\longrightarrow} \int_0^t \sigma_s^2 d s.
\end{equation}
Due to the nature of these two estimators, their difference allows capturing the jump component of the quadratic variation
$R V-B V \stackrel{p}{\rightarrow} \sum_{0 \leq s \leq t} \Delta p_s^2$

The RV estimator can be split into two signed realized semivariances, $RV = RV^+ + RV^-$, that respectively capture variation due to positive and negative returns
\begin{align}\label{Eq:RS}
& RV^{+}=\sum r_i^2 I\left[r_i>0\right] \stackrel{p}{\longrightarrow} 1 / 2 \int_0^t \sigma_s^2 d s+\sum_{0<s<=t}\left(\Delta p_s^2\right) I\left[\Delta p_s>0\right] \nonumber \\
& RV^{-}=\sum r_i^2 I\left[r_i<0\right] \stackrel{p}{\longrightarrow} 1 / 2 \int_0^t \sigma_s^2 d s+\sum_{0<s<=t}\left(\Delta p_s^2\right) I\left[\Delta p_s<0\right].
\end{align}
These measures have been shown to include variations of both the continuous and the jump part of Eq. \ref{Eq:qv}. The first term is equal for both semivariances and not decomposable into a signed measure. It follows then that jump variation is defined as 
\begin{equation}
\Delta J^2  \equiv RV^{+}-RV^{-}  \stackrel{p}{\rightarrow} \sum_{0 \leq s \leq l} \Delta p_s^2 I\left\{\Delta p_s>0\right\}-\sum_{0 \leq s \leq l} \Delta p_s^2 I\left\{\Delta p_s<0\right\}
\end{equation}
The difference eliminates the common integrated variance term and is positive when positive shocks are prevalent in a day and negative if adverse shocks dominate a day. Using the signed jumps component instead of the whole jump component, RV - BV, gives the advantage of separately studying the impact of significant positive and negative price movements on future volatility. \cite{merton1976option} first introduce the idea of a jump component as a jump-diffusion process where large variations of prices occur at a discrete time together with small continuous movements.

Each of these variance estimators is obtained from a series of equally-spaced observed prices $p_0, p_1, \ldots, p_N$ that span through the time horizon of the analysis. Setting $N$ as the number of intraday observations available daily, the RV estimator is the sum of intraday squared returns following Eq. \ref{Eq:RV}, as well as realized semivariances, $RV^{+}$and $RV^{-}$, follows Eq. \ref{Eq:RS}. 

The estimator for the bipower variation BV in Eq. \ref{Eq:BV} is obtained by averaging multiple skip versions of the same estimator defined as 
\begin{equation}
    BV=\frac{\pi}{2} \sum_{i=q+2}^n\left|r_i\right|\left|r_{i-1-q}\right| 
\end{equation}
where the usual $BV$ estimator is obtained when $q=0$. We obtain the BV estimator by averaging skip-0 through skip-4 of the same estimator as an important correction for the bipower variation estimator.

The purpose of computing the estimators described in this Section is to obtain a daily proxy for the quadratic variation of the log-return prices process from high-frequency data. By disentangling the different components of the quadratic variation into their relative estimators, we can empirically analyze the effect of different volatility drivers in the cryptocurrency market, such as the signed part of the realized variance and the jump components. 

\subsection{Data Collection}\label{Subsec:data}
We obtained the 5-minute level prices of the 100 most capitalized cryptocurrencies\footnote{The list is available \href{https://coinmarketcap.com/historical/20221013/}{here}.} from the Binance API (Application Programming Interface) \footnote{https://www.binance.com/en/binance-api} at the time of writing, such that our sample covers more than 95\% of the cryptocurrency market capitalization. From the 100 cryptocurrencies, we exclude stablecoins\footnote{Stablecoins are cryptocurrencies designed to maintain a stable value relative to a specific asset, such as the US dollar or a basket of assets. The value of stablecoins is backed by these assets, which are held in reserve, and they aim to minimize price volatility compared to other cryptocurrencies like Bitcoin and Ethereum \citep{fiedler2023stablecoins}.} from our analysis since we are interested in volatility fluctuations that are not expected within that category of digital assets, remaining with 87 entities. Then, we queried from Polygon\footnote{https://polygon.io/} the stock prices of the 42 companies included in the Nasdaq technology index with the same granularity. To compare the two asset classes, we consider the historical period between the year 2020 and year 2022 (up to October). Many of the most highly capitalized cryptocurrencies at the time of this analysis began trading after 2020, making the sample representative of the current state of the market.

The amount of 5-minute level observation daily varies among cryptocurrencies, continuously traded throughout the day, and stock, which follows the market hours. This results in a different amount of high-frequency observation each day in our dataset, 87 for cryptocurrencies and 79 for stocks. We compute the estimators using a different number of observations to properly reflect the realized variance within a given day for both asset classes. The entities in the cryptocurrency cross-section have around four times the available observation compared to equities. They are, therefore, subject to be more volatile by construction since each entity has more recorded fluctuations. However, this difference does not impact how we construct $n$-days volatility estimators that follow the same process for both asset classes. 

\ref{tab:descriptive_stats} contains the average values, the standard deviations, and some quantiles of the distribution of the computed estimators. The table is divided to provide the same information for the cross-sections of entities included in our analysis. The included estimators, realized variance (RV ), bipower variation (BV ), positive and negative semivariance (RV- and RV+), jump variation (SJV), and signed jump variation (SJV- and SJV+), are those involved in the autoregressive models outlined in the next Section. The descriptive statistics show a higher order of magnitude for the cryptocurrency estimators than those related to equities. This effect indeed points to the continuous trading under which cryptocurrencies are subjected due to how the estimator is constructed. However, the difference in scale is also due to the more significant volatility shocks that are known to affect such a market.

\ref{tab:correlation} contains the two correlation matrices for the computed estimators. For both the cross-sections, the RV estimator is highly correlated to its signed components and the continuous volatility estimator BV, as expected by construction. On the contrary, the SJV estimators and their signed components are not strongly correlated with the other estimators for the equity cross-section. At the same time, such a correlation tends to be higher in the case of cryptocurrencies. Such results potentially signal a stronger influence of the jump part to explain future cryptocurrency volatility.

\begin{table}[htb]
    \centering
    \begin{tabular}{llrrrrrrr}
    \toprule
    {} & {} &   Mean &    Std &     5\% &    25\% &    50\% &    75\% &    95\% \\
    \midrule
    \multirow{7}{*}{\rotatebox[origin=c]{90}{Crypto}} & RV   &  0.459 &  2.881 &  0.036 &  0.112 &  0.222 &  0.448 &  1.397 \\
    & BV   &  0.404 &  2.510 &  0.030 &  0.095 &  0.191 &  0.391 &  1.244 \\
    & RV-  &  0.228 &  1.806 &  0.017 &  0.055 &  0.110 &  0.222 &  0.679 \\
    & RV+  &  0.230 &  1.124 &  0.017 &  0.053 &  0.106 &  0.223 &  0.731 \\
    & SJV  &  0.002 &  0.869 & -0.124 & -0.024 & -0.001 &  0.023 &  0.164 \\
    & SJV- & -0.037 &  0.824 & -0.124 & -0.024 & -0.001 &  0.0002 &  0.0005 \\
    & SJV+ &  0.039 &  0.270 & -0.012 &  0.001 & 0.003 &  0.023 &  0.164 \\
    \midrule
    \multirow{7}{*}{\rotatebox[origin=c]{90}{Equity}} & RV   &  0.069 &  0.284 &  0.002 &  0.008 &  0.010 &  0.053 &  0.284 \\
    & BV   &  0.025 &  0.070 &  0.022 &  0.061 &  0.001 &  0.025 &  0.110 \\
    & RV-  &  0.035 &  0.157 &  0.016 &  0.023 &  0.003 &  0.024 &  0.141 \\
    & RV+  &  0.035 &  0.153 &  0.012 &  0.026 &  0.003 &  0.024 &  0.147 \\
    & SJV  &  0.001 &  0.123 & -0.046 & -0.001 &  0.0005 &  0.001 &  0.053 \\
    & SJV- & -0.012 &  0.090 & -0.046 & -0.001 &  0.0004 &  0.001 &  0.003 \\
    & SJV+ &  0.012 &  0.083 &  0.031 &  0.001 &  0.0002 &  0.004 &  0.053 \\
    \bottomrule
    \end{tabular}
    \caption{Descriptive statistics expressed in percentage of the constructed variance estimators for cryptocurrencies and equities.}
    \label{tab:descriptive_stats}
\end{table}

\begin{table}[htb]
    \centering
    \begin{tabular}{llrrrrrrr}
    \toprule
    {} & {} &       RV &       BV &      RV- &      RV+ &      SJV &     SJV- &     SJV+ \\
    \midrule
    \multirow{7}{*}{\rotatebox[origin=c]{90}{Crypto}} & RV   &  1.000 &  0.995 &  0.989 &  0.972 & -0.798 & -0.901 &  0.179 \\
    & BV   &  0.995 &  - &  0.984 &  0.968 & -0.793 & -0.878 &  0.128 \\
    & RV-  &  0.989 &  0.984 &  - &  0.928 & -0.877 & -0.947 &  0.067 \\
    & RV+  &  0.972 &  0.968 &  0.928 &  - & -0.637 & -0.787 &  0.352 \\
    & SJV  & -0.798 & -0.793 & -0.877 & -0.637 &  - &  0.950 &  0.317 \\
    & SJV- & -0.901 & -0.878 & -0.947 & -0.787 &  0.950 &  - &  0.006 \\
    & SJV+ &  0.179 &  0.128 &  0.067 &  0.352 &  0.317 &  0.006 &  - \\
    \midrule
    \multirow{7}{*}{\rotatebox[origin=c]{90}{Equity}} & RV   &  - &  0.819 &  0.919 &  0.914 & -0.036 & -0.539 &  0.527 \\
    & BV   &  0.819 &  - &  0.784 &  0.719 & -0.106 & -0.424 &  0.298 \\
    & RV-  &  0.919 &  0.784 &  - &  0.684 & -0.426 & -0.778 &  0.206 \\
    & RV+  &  0.914 &  0.719 &  0.682 &  - &  0.370 & -0.203 &  0.767 \\
    & SJV  & -0.036 & -0.106 & -0.426 &  0.370 &  - &  0.738 &  0.688 \\
    & SJV- & -0.539 & -0.424 & -0.778 & -0.203 &  0.738 &  - &  0.019 \\
    & SJV+ &  0.527 &  0.298 &  0.206 &  0.767 &  0.688 &  0.01930 &  - \\
    \bottomrule
    \end{tabular}
    \caption{Linear correlation matrices of the constructed variance estimators for cryptocurrencies and equities.}
    \label{tab:correlation}
\end{table}

\subsection{Models}\label{Subsec:models}
To study the driving factor of the volatility within the two asset classes, we consider a set of different model specifications that estimate future volatility using the information the computed volatility estimators provided. All the model specifications are estimated in-sample since our scope is to obtain insights regarding the driving factor of the volatility process. The forecasting problem is not treated in this study, although the results can be beneficial to improve modeling for that purpose.

The reference model specification is the heterogeneous autoregression (HAR) specification \citep{corsi2009simple}, which is a high-order autoregressive model specified as 
\begin{equation}\label{Eq:refmodel}
RV_{h, t+h}=  \mu+\phi_d RV_t+\phi_w\left(\frac{1}{4} \sum_{i=1}^4 RV_{t-i}\right)+\phi_m\left(\frac{1}{17} \sum_{i=5}^{21} RV_{t-i}\right) +\epsilon_{t+h}.
\end{equation}
The model regresses the future realized variance estimator $t+h$ on three past terms of the same estimator: the first lag, the average between the second and the fifth lag, and the average between the sixth and the twenty-second lag. These three regressors intuitively correspond to an estimation of the volatility on the previous day and the average volatility over the past trading week and trading month. The model is said to be rotated since each regressor has no overlapped observation, i.e., the average of the past trading week does not include the observation of the previous day since another regressor already represents it. We estimate the reference specification and each of its modifications at four different horizons, corresponding to one day, a trading week (5 days ahead), a trading month (22 days ahead), and three trading months (66 days ahead) in the future. This model specification looks for the effects of three representative market agents operating at those three different frequencies.

We compute the realized variance estimator described in \ref{Subsec:estimators} using the 5-minute level data in \ref{Subsec:data} for each of the cryptocurrencies. Once we estimate these volatility proxies, it is straightforward to derive the regressor of the reference model in Eq. \ref{Eq:refmodel} by lagging or averaging over the information according to the specification. The HAR model's main benefit is that it allows for incorporating multiple time scales, which is particularly important because volatility exhibits specific patterns at different frequencies. The HAR model is an extension of the Autoregressive Conditional Heteroskedasticity (ARCH) model, which assumes that the variance of a time series is a function of its past squared residuals and can capture the long-memory properties of financial volatility. Lastly, it is a relatively easy-to-interpret model, which makes it suitable for descriptive analyses of linear volatility patterns. However, the reduced complexity can cause some information regarding nonlinear volatility behaviors to be missing.

Starting from the reference model and following the framework of \cite{patton2015good}, we estimate various HAR specifications, each accounting for specific traits and stylized facts of the volatility dynamics. Each model specification analyzes the effect of signed information, such as the positive and negative part of the realized variance and the jump components of the price process, in improving modeling and explainability of the global cryptocurrency market volatility. All the model specifications tested are derived from the reference model HAR-RV, which we rewrite from \ref{Eq:refmodel} in a compact form as
 
\begin{equation}\label{Eq:refcompact}
    RV_{h,t+h} = \mu + \phi_d RV_t  + \phi_w RV_{w,t} + \phi_m RV_{m,t} + \epsilon_{t+h}
\end{equation}

Recalling the definition of RV as the sum of squared returns, one test for the effect of signed returns through the realized semivariances in \ref{Eq:semirv}. The HAR-semiRV specification corresponds to
\begin{equation}\label{Eq:semirv}
     RV_{h,t+h} = \mu + \phi^+_d RV^+_t + \phi^-_d RV^-_t + \phi_w RV_{w,t} + \phi_m RV_{m,t} + \epsilon_{t+h}
\end{equation}
when the most recent information is decomposed into the signed realized semivariances. The \ref{tab:fullsemiHAR} in the appendix also estimates a similar model specification when the three regressors are all decomposed into the signed components. Such an autoregressive model detects the effect of the market's direction on future volatility, describing the entity of the leverage effect at the intraday level. As previously discussed, the granularity of the data collected to obtain the volatility estimators defines the amount of information contained in the regressor for the model specification. Therefore, the leverage effect detected daily could differ from that detected on a more fine-grained scale.

To also account for a daily leverage effect, in addition to the intraday leverage effect captured by the semivariances, we also estimate the model specification in Eq. \ref{Eq:refcompact} and \ref{Eq:semirv} by including a supplementary interaction term $RV_t I_{r_t<0}$:
\begin{align}\label{Eq:leveffect}
    RV_{h,t+h} &= \mu + \phi_d RV_t +  \gamma RV_t I_{r_t<0} + \phi_w RV_{w,t} + \phi_m RV_{m,t} + \epsilon_{t+h}   \\
    RV_{h,t+h} &= \mu +  \phi^+_d RV^+_t + \phi^-_d RV^-_t + \gamma RV_t I_{r_t<0} + \phi_w RV_{w,t} + \phi_m RV_{m,t} + \epsilon_{t+h}.
\end{align}
which is nonzero only when the daily return of the asset under study is negative. This term reflects the definition of the leverage effect from the traditional financial literature, where a negative price variation increases the volatility more than a positive one. Since the interaction term is based on daily information, while realized variance and semivariances are based on 5-minute level prices, we expect to find a different effect in the results.

An additional couple of tested model specifications disentangle the most recent information into the continuous and the jump part of the volatility. We obtain two variants, where the second one also decomposes the jump part to account for its signed effect.
\begin{align}
    RV_{h,t+h} &= \mu + \phi_j \Delta J^2_t + \phi_c BV_t + \phi_w RV_{w,t} + \phi_m RV_{m,t} + \epsilon_{t+h}  \\
    RV_{h,t+h} &= \mu  + \phi_j+ \Delta J^{2+}_t +  \phi_j- \Delta J^{2-}_t + \phi_c BV_t + \phi_w RV_{w,t} + \phi_m RV_{m,t} + \epsilon_{t+h}.
\end{align}
It is possible to calculate the jump parts of the volatility estimators from Eq. \ref{Eq:RV} and \ref{Eq:BV} because the BV estimators converge in the limit to the continuous component of the quadratic variation of the price process. Including the jump component into a model specification is extremely important in markets characterized by a lack of regulation and liquidity, as is the case for cryptocurrencies. \ref{tab:descriptive_stats} highlights the different scale of magnitude of the estimated realized variance for the cryptocurrency cross-section compared to the equity one as a signal of more frequent large shifts in the price variation.

For each model specification described in this section, we estimate a panel regression to capture the average contribution of each entity in a specific asset class and explain the volatility dynamics for the considered group. Therefore, the panel regression is carried out separately for cryptocurrencies and equities for a thoughtful comparison. Each panel HAR is estimated through a weighted least squares (WLS) method to account for frequent regime shifting retrieved in the dynamics of our time series. The WLS estimation is based on a preliminary Ordinary Least Squares (OLS) estimation, which is necessary to construct weights as the inverse of the fitted residual from such an OLS model. As a further analysis, the same estimation sequence is repeated on different time windows to capture the effect of different market regimes and separately on each component of the cross-sections to disentangle the aggregated effect at the entity level and validate the findings. When estimating the panel regressions, we adopt the Newey-West Heteroskedasticity and Autocorrelation Consistent (HAC) estimators. This approach is particularly useful for addressing the potential issues of autocorrelation and heteroskedasticity in panel data, which are common in financial time series data. The Newey-West HAC estimators allow us to produce standard errors that are robust to these issues, thereby ensuring that our inference is more reliable.

\section{Empirical Results}\label{Sec:result}
This Section provides the different results of our empirical analysis where we comment on the estimation results of various model specifications described in \ref{Subsec:models}. The estimation procedure remains the same as described, even when we restrict the time window of observation used for the estimation and when we estimate the model specifications individually over each entity of a specific cross-section instead of as a panel.
\subsection{Impact of Signed Returns}\label{Sec:signret}
As a first step in the empirical analysis of cryptocurrency volatility, we estimate the reference specification that accounts for three different lags of the estimated volatility observed in the past. The results on the reference model described in Eq. \ref{Eq:refcompact} are instrumental in describing past signed returns' impact on future volatility.

\ref{tab:reference_HAR_semiHAR} presents the results for the reference specification of the panel HAR and its variant where the first lag of the volatility is decomposed into its signed components. Both model specifications are separately estimated for the cryptocurrency and the equity entities, as shown in the table.


The first result to highlight is the relative amount of volatility persistence, defined as $\phi_d + \phi_w + \phi_m = 1$ in case of highest persistence, for the first lag of the reference specification of cryptocurrency cross-section at the horizon $t+1$ and $t+5$. This effect is remarkably missing in crypto, where the relative contribution to the volatility persistence is low for all the estimation horizons. This result aligns with previous findings on cryptocurrency volatility persistence until 2018 \citep{katsiampa2019high,charfeddine2019shocks,abakah2020volatility,caporale2018persistence}. Higher volatility persistence suggests that the market's reaction to new information entering the financial system is not instantaneous but occurs progressively over time. Consequently, historical price fluctuations can serve as indicators for forecasting future price movements, offering a counterpoint against the Efficient Market Hypothesis (EMH), indicating that the cryptocurrency markets may not always reflect all available information immediately and efficiently \citep{cross2021returns}. This preliminary observation serves as a starting point for a more comprehensive discussion on the Adaptive Market Hypothesis (AMH) \citep{lo2004adaptive}, which we explore in detail in \ref{Sec:connections}. The AMH provides a framework for understanding the evolving efficiency of the cryptocurrency market, highlighting how its unique ecosystem characteristics and the interplay of various market participants contribute to its dynamic volatility patterns.

The role of recent information, as measured by $\phi_d$, is prevalent in the crypto cross-section, while it tends to be more spread across different lags for equity. This result signals the temporary nature of the volatility in the cryptocurrency market, whose future values are not largely correlated with distant observations in the past. A similar effect regarding the importance of recent information to estimate volatility is captured by looking at the diminishing role of $\phi_d$ as the horizon increases. Although more pronounced for cryptocurrencies, such an effect is shared among the two asset classes for the time period considered.

We observe that the decomposition of the most recent information on volatility into its signed counterparts adds information to the analysis since  $\phi^+_d \neq \phi^-_d \neq \phi_d$ for both asset classes. The effect of the positive returns is always significant and dominates the impact of the negative returns. It is important to notice that, when statistically significant, the negative part of the volatility has the opposite effect on cryptocurrencies compared to what is observed for equities. In general, we observe that the role of the negative volatility is negligible for cryptocurrencies as a signal of the missing impact of negative shocks of the market on future volatility, contrary to what is usual for other asset classes. Such a result implies the absence of the leverage effect for cryptocurrencies when using high-frequency data. The decreasing impact of recent information is also present when using the signed past returns for the most recent lag. Those effects are present when the RV estimator is decomposed into the signed components at various lags. The \ref{tab:fullsemiHAR} in the appendix provides the estimated coefficients for such model specifications.

The results obtained at a high-frequency level for the cryptocurrency cross-section highlight an absence of the leverage effect in the traditional sense, where a negative return impacts future volatility more than a positive one. Such behavior challenges the notion of market efficiency and the traded price as an aggregator of all the available information. The hypothesis of rational expectation in front of a negative return falls in the case of cryptocurrency since it is a positive movement of the market that increases the volatility. The market participants are not necessarily rational in the cryptocurrency market, as the massive presence of retail investors exacerbates idiosyncratic volatility in unstable market conditions \citep{low2019cryptocurrency,ozdamar2022retail,wen2019retail}. The next subsection elaborates more on the leverage effect aspect by comparing the high-frequency effect with the daily one.

Another motivation for retrieving unexpected volatility patterns in the cryptocurrency market compared to the equity one is the different mechanisms and set of technologies the ecosystem builds upon (see \cite{xu2019systematic} for a review of the Blockchain technology). Even though the level of on-chain activity\footnote{
On-chain market activity refers to transactions recorded on the blockchain, while off-chain market activity refers to transactions that occur outside of the blockchain, typically facilitated by third-party services such as centralized exchanges or wallets.} is still a fraction of the volume of off-chain trades that happens on centralized exchanges, it is in the process of growing and represents an element of novelty compared to traditional finance. Our analysis focuses on off-chain data for comparison with the equity market, but substantial activity on a decentralized platform makes the market less reactive and fractionated \citep{aspris2021decentralized}. Indeed, traders can often find arbitrages between centralized and decentralized exchanges. Hence, the cryptocurrency market is slower than the equity one at transferring information into the quoted prices due to these two concurrent layers of activity.

\begin{singlespacing}
\begin{table}[H]
    \caption*{\small{$RV_{h,t+h} = \mu + \phi_d RV_t +   \phi^-_d RV^-_t + \phi^+_d RV^+_t + \phi_w RV_{w,t} + \phi_m RV_{m,t} + \epsilon_{t+h}$}}
\centering
\begin{tabular}{ccccccc} 
\toprule
        &       & $\mathbf{\phi_d}$ &  $\mathbf{\phi^-_d}$ &  $\mathbf{\phi^+_d}$  & $\mathbf{\phi_w}$ & $\mathbf{\phi_m}$  \\
\midrule
 \multirow{16}{*}{\rotatebox[origin=c]{90}{Crypto}} & \multirow{4}{*}{\text{t+1}} &       0.554\threeS &                - &       - &                          0.006 &       0.056\threeS \\
 
     &                            &       {\footnotesize{(6.726)}} &             - &       - &                     {\footnotesize{{(1.573)}}} &      {\footnotesize{(4.025)}} \\[1.0ex]
                                 
&  &         - &                -0.142\oneS &       1.262\threeS &                         0.007\oneS &       0.036\threeS \\
  
     &                            &    -    &                         {\footnotesize{(-1.694)}}&       {\footnotesize{(10.183)}} &                     {\footnotesize{{(1.774)}}} &      {\footnotesize{(4.764)}} \\[1.0ex]

\cmidrule{3-7}
& \multirow{4}{*}{\text{t+5}} &       0.271\threeS &                - &            - &                     0.063\threeS &       0.051\threeS \\
 
     &                           &       {\footnotesize{(4.834)}} &                           -&       - &                     {\footnotesize{{(2.773)}}} &      {\footnotesize{(3.585)}} \\[1.0ex]
                                
 & &         - &                0.078 &       0.543\threeS &                         0.054\threeS &       0.032\threeS \\
  
      &                           &    -    &                         {\footnotesize{(0.258)}}&       {\footnotesize{(2.313)}} &                     {\footnotesize{{(2.4)}}} &      {\footnotesize{(1.536)}} \\[1.0ex]

\cmidrule{3-7}
 &\multirow{4}{*}{\text{t+22}} &       0.076\threeS &                - &             - &                     0.031\threeS &       0.049\threeS \\
 
     &                             &       {\footnotesize{(3.551)}} &                            -&       - &                     {\footnotesize{{(2.681)}}} &      {\footnotesize{(4.089)}} \\[1.0ex]
                                  
 & &         - &                -0.065 &      0.24\threeS &                         0.028\threeS &       0.043\threeS \\
  
    &                             &    -    &                         {\footnotesize{(-0.946)}}&       {\footnotesize{(4.33)}} &                     {\footnotesize{{(2.592)}}} &      {\footnotesize{(3.764)}} \\[1.0ex]

\cmidrule{3-7}
 &\multirow{4}{*}{\text{t+66}} &       0.025\twoS &                - &             - &                     0.006 &       -0.012\threeS \\
 
    &                              &       {\footnotesize{(2.335)}} &                           -&       - &                     {\footnotesize{{(1.293)}}} &      {\footnotesize{(-2.676)}} \\[1.0ex]
                                  
 & &         - &                -0.069\threeS &       0.145\threeS &                        0.006\oneS &       -0.016\threeS \\
  
     &                            &    -    &                         {\footnotesize{(-4.7734)}}&       {\footnotesize{(8.392)}} &                     {\footnotesize{{(1.174)}}} &      {\footnotesize{(-3.814)}} \\[1.0ex]
                                  
\midrule
  \multirow{16}{*}{\rotatebox[origin=c]{90}{Equity}} & \multirow{4}{*}{\text{t+1}} &   0.187\threeS &                - &       - &                          0.182\threeS &       0.165\threeS \\
 
     &   &                               {\footnotesize{(7.553)}} &             - &       - &                     {\footnotesize{{(7.419)}}} &      {\footnotesize{(3.081)}} \\[1.0ex]
                                 
 &   &       - &                0.075\twoS &       0.305\threeS &                         0.182\threeS &       0.163\threeS \\
  
  &    &                              -    &                         {\footnotesize{(2.276)}}&       {\footnotesize{(5.339)}} &                     {\footnotesize{{(7.527)}}} &      {\footnotesize{(3.051)}} \\[1.0ex]

\cmidrule{3-7}
   &   \multirow{4}{*}{\text{t+5}}   & 0.089\threeS &                - &            - &                     0.183\threeS &       0.162\threeS \\
 
   &  &                                  {\footnotesize{(7.95)}} &                           -&       - &                     {\footnotesize{{(12.1)}}} &      {\footnotesize{(6.211)}} \\[1.0ex]
                                
    & &      - &                0.055\twoS &       0.125\threeS &                         0.184\threeS &       0.162\threeS \\
  
     &  &                              -    &                         {\footnotesize{(2.45)}}&       {\footnotesize{(5.011)}} &                     {\footnotesize{{(12.086)}}} &      {\footnotesize{(6.197)}} \\[1.0ex]

\cmidrule{3-7}
  & \multirow{4}{*}{\text{t+22}}   &  0.048\threeS &                - &             - &                     0.108\threeS &       0.169\threeS \\
 
     &  &                                  {\footnotesize{(6.42)}} &                            -&       - &                     {\footnotesize{{(10.359)}}} &      {\footnotesize{(12.428)}} \\[1.0ex]
                                  
     & &     - &                0.01\threeS &      0.088\threeS &                         0.108\threeS &       0.169\threeS \\
  
    &  &                               -    &                         {\footnotesize{(2.811)}}&       {\footnotesize{(5.513)}} &                     {\footnotesize{{(10.393)}}} &      {\footnotesize{(12.428)}} \\[1.0ex]

\cmidrule{3-7}
 &  \multirow{4}{*}{\text{t+66}}   &  0.028\threeS &                - &             - &                     0.08\threeS &       0.217\threeS \\
 
     & &                                   {\footnotesize{(5.967)}} &                           -&       - &                     {\footnotesize{{(11.21)}}} &      {\footnotesize{(23.229)}} \\[1.0ex]
                                  
& & - &                0.01\twoS &      0.047\threeS &                         0.08\threeS &       0.216\threeS \\
  
  &       &                            -    &                         {\footnotesize{(2.036)}}&       {\footnotesize{(4.819)}} &                     {\footnotesize{{(11.233)}}} &      {\footnotesize{(23.254)}} \\[1.0ex]
\bottomrule
\addlinespace[1ex]
\multicolumn{3}{l}{\textsuperscript{***}$p<0.01$, 
  \textsuperscript{**}$p<0.05$, 
  \textsuperscript{*}$p<0.1$}
\end{tabular}                        
    \vspace{1em}
    \caption{Estimation results for HAR-RV and HAR-semiRV model specifications. Results are displayed for the cryptocurrency and the equity panel over the four different horizons. T-stats are in parentheses. The model equation on top of the table reflects all the parameters common to the specification reported in this table. The total number of observations available considering all the entities used for the panel estimate is 47787 for crypto and 17111 for equity}
    \label{tab:reference_HAR_semiHAR}
\end{table}
\end{singlespacing}
\subsection{Daily Leverage Effect}\label{Subsec:lev}

In this section, we introduce a daily leverage effect into the reference specification estimated in the previous section. The leverage effect is a known stylized fact of financial returns that exhibits an asymmetric impact of financial returns on the future volatility of the assets. The financial econometrics literature has extensively investigated such relationships at a daily frequency, usually through Generalized AutoRegressive Conditional Heteroskedasticity, GARCH-like, \citep{bollerslev1987conditionally} models.

We add a daily leverage effect component, $\gamma RV_t I_{r_t<0}$, according to Eq. \ref{Eq:leveffect}. This model specification allows us to evaluate the contribution of the decomposition of the realized variance at a high-frequency level, adding value beyond the usual daily effect. Note that the daily leverage effect is modeled using a lagged squared return interacted with an indicator for negative returns, similar to an asymmetric GARCH-like model as the GJR-GARCH \citep{glosten1993relation}. \ref{tab:HAR_leverage} shows the estimation results with the same structure proposed in the previous section. If the addition of volatility estimator obtained from high-frequency data did not add additional information, we would expect the estimation results to show $\phi^+_d = \phi^-_d = \phi_d$ and the daily leverage parameter $\gamma$ as statistically significant. The results show that the positive part of the most recent realized variance lag is always statistically significant at any time horizon for both asset classes. In contrast, the negative part does not have enough statistical evidence. The sign of the daily leverage effect interaction is opposite for the cryptocurrency panel compared to the equity panel, being negative for the former and positive for the latter. The inversion of the sign also holds when the daily leverage component is included in the reference specification. These results are consistent with the leverage effect inversion in crypto due to the diminishing effect of a negative shock on the realized variance.

This lack of statistical significance serves as an indicator of an inversion in the leverage effect within the cryptocurrency market. The analysis demonstrates that the expected negative influence typically associated with downturns fails to materialize. Instead, this anomaly underscores the unique dynamics, suggesting that the traditional leverage effect, as observed in other financial markets, does not apply in the same manner to cryptocurrencies. While the results of \ref{Sec:signret} show statistically insignificant measures for the negative part of the realized variance only for some horizons, the results hold even more when the daily leverage component is included. This inversion is a critical insight, highlighting the distinct market responses to positive versus negative shocks in the realm of cryptocurrencies. The daily leverage aggregation provided by the $\gamma RV_t I_{r_t<0}$ factor cancels out the effects captured by the high-frequency semivariance estimator, especially if the leverage effect operates differently within shorter time frames and is more prevalent at the daily level as already explored by \cite{baur2018asymmetric,brini2022assessing}.

The reversed asymmetric effect of high-frequency cryptocurrency returns indicates that investors exploit market crashes as a buying opportunity to enter at a discount. Such behavior connects to FoMo as a motivator for investment behavior in the cryptocurrency market. \cite{gerrans2023fear} retrieves a larger association between FoMO and crypto ownership than equity ownership for young age groups more likely to trade crypto assets. The concept of FoMo among retail investors has been explored by \cite{argan2023role} to observe the relationship with investment engagement and by \cite{park2023fear} to analyze the relationship with market stability. \cite{kaur2023all} observes the FoMo effect in crypto investment decisions as a driving factor for the herding effect in investment decisions, while \cite{delfabbro2021psychology} delve into the psychological traits of cryptocurrency trading, among which FoMo represents one of the factors. \cite{king2021herding} elaborates on the effect of herding behavior on the trading activity for cryptocurrencies. We argue that the fear of missing out on potential profits when others are succeeding is a potential cause for the leverage effect inversion retrieved by the econometric analysis. Specifically, the market is relatively more active, hence more volatile, when returns are positive and spark excitement in the crowd that wants to enter this new investment opportunity, rather than when returns are negative and sell-offs happen. Psychology-driven behaviors can cause retail investors to enter the market without clearly understanding the asset class, viewing it as a new opportunity with high potential \citep{pursiainen2022retail}. Indeed, it is known that the retail investors' presence is larger in crypto than equity due to the nascent nature of the former and other peculiar characteristics such as the relatively low entry barrier, the potential to disrupt the conventional financial system and the underlying Blockchain technology that powers the ecosystem.

\begin{singlespacing}
\begin{table}[H]
    \caption*{\small{$RV_{h,t+h} = \mu + \phi_d RV_t +   \phi^-_d RV^-_t + \phi^+_d RV^+_t + \gamma RV_t I_{r_t<0} + \phi_w RV_{w,t} + \phi_m RV_{m,t} + \epsilon_{t+h}$}}
\centering
\begin{tabular}{cccccccc} 
\toprule
        &       & $\mathbf{\phi_d}$ &  $\mathbf{\phi^-_d}$ &  $\mathbf{\phi^+_d}$  &  $\mathbf{\gamma}$  & $\mathbf{\phi_w}$ & $\mathbf{\phi_m}$  \\
\midrule
 \multirow{16}{*}{\rotatebox[origin=c]{90}{Crypto}} & \multirow{4}{*}{\text{t+1}} &       0.632\threeS &                - &       - &  -0.126 &                        0.006 &       0.054\threeS \\
 
     &                            &       {\footnotesize{(16.698)}} &             - &       - & {\footnotesize{(-1.444)}} &                    {\footnotesize{{(1.455)}}} &      {\footnotesize{(4.163)}} \\[1.0ex]
                                 
  & & - &         0.298 &                0.661\threeS &       -0.261\threeS &       0.026 &                  0.127\threeS        \\
&     &               -             &    {\footnotesize{(1.436)}}    &                         {\footnotesize{(3.355)}}&       {\footnotesize{(-2.755)}} &  {\footnotesize{(1.551)}} &                   {\footnotesize{{(4.627)}}}  \\[1.0ex]

\cmidrule{3-8}
& \multirow{4}{*}{\text{t+5}} &       0.32\threeS &                - &            - &        -0.053 &             0.056\threeS &       0.047\threeS \\
 
     &                           &       {\footnotesize{(12.709)}} &                           -&       - &      {\footnotesize{(-0.929)}} &              {\footnotesize{{(2.645)}}} &      {\footnotesize{(3.445)}} \\[1.0ex]
                                
  & & - &         0.314 &                0.151 &       0.026 &       0.084\threeS &                  0.036\threeS        \\
&     &               -             &    {\footnotesize{(0.712)}}    &                         {\footnotesize{(0.0363)}}&       {\footnotesize{(0.247)}} &  {\footnotesize{(3.072)}} &                   {\footnotesize{{(1.485)}}}  \\[1.0ex]

\cmidrule{3-8}
 &\multirow{4}{*}{\text{t+22}} &       0.145\threeS &                - &             - &          -0.083\threeS &           0.026\twoS &       0.043\threeS \\
 
     &                             &       {\footnotesize{(13.745)}} &                            -&       - &           {\footnotesize{(-4.183)}} &          {\footnotesize{{(2.534)}}} &      {\footnotesize{(3.777)}} \\[1.0ex]
                                  
  & & - &         0.07 &                0.196\threeS &       -0.076\threeS &       0.029\threeS &                  0.043\threeS        \\
&     &               -             &    {\footnotesize{(0.822)}}    &                         {\footnotesize{(3.254)}}&       {\footnotesize{(-3.852)}} &  {\footnotesize{(2.618)}} &                   {\footnotesize{{(3.815)}}}  \\[1.0ex]

\cmidrule{3-8}
 &\multirow{4}{*}{\text{t+66}} &       0.084\threeS &                - &             - &        -0.068\threeS &             0.004 &       -0.016\threeS \\
 
    &                              &       {\footnotesize{(13.376)}} &              - &             -&       {\footnotesize{(-7.32)}} &                     {\footnotesize{{(0.917)}}} &      {\footnotesize{(-4.033)}} \\[1.0ex]
                                  
  & & - &         0.298 &                0.661\threeS &       -0.261\threeS &       0.026 &                  0.127\threeS        \\
&     &               -             &    {\footnotesize{(1.436)}}    &                         {\footnotesize{(3.355)}}&       {\footnotesize{(-2.755)}} &  {\footnotesize{(1.551)}} &                   {\footnotesize{{(4.627)}}}  \\[1.0ex]
                                  
\midrule
  \multirow{16}{*}{\rotatebox[origin=c]{90}{Equity}} & \multirow{4}{*}{\text{t+1}} &   0.138\threeS &          - &      - &       0.09\twoS &                          0.183\threeS &       0.166\threeS \\
 
     &   &                               {\footnotesize{(4.403)}} &             - &       - &           {\footnotesize{(1.988)}} &          {\footnotesize{{(7.471)}}} &      {\footnotesize{(3.137)}} \\[1.0ex]
                                 
  & & - &         -0.012 &                0.266\threeS &       0.115\threeS &       0.184\threeS &                  0.165\threeS        \\
&     &               -             &    {\footnotesize{(-0.319)}}    &                         {\footnotesize{(4.578)}}&       {\footnotesize{(2.721)}} &  {\footnotesize{(7.635)}} &                   {\footnotesize{{(3.13)}}}  \\[1.0ex]

\cmidrule{3-8}
   &   \multirow{4}{*}{\text{t+5}}   & 0.064\threeS &                - &            - &        0.047\twoS &             0.183\threeS &       0.161\threeS \\
 
&     &               {\footnotesize{(5.443)}}              &   -    &          -     &          {\footnotesize{(2.417)}}&       {\footnotesize{(12.113)}} &  {\footnotesize{(6.197)}}   \\[1.0ex]
                                
  & & - &         0.012 &                0.107\threeS &       0.055\twoS &       0.184\threeS &                  0.161\threeS        \\
&     &               -             &    {\footnotesize{(0.526)}}    &                         {\footnotesize{(4.139)}}&       {\footnotesize{(2.193)}} &  {\footnotesize{(12.105)}} &                   {\footnotesize{{(6.183)}}}  \\[1.0ex]

\cmidrule{3-8}
  & \multirow{4}{*}{\text{t+22}}   &  0.036\threeS &                - &             - &         0.021\oneS &            0.108\threeS &       0.169\threeS \\
 
   &  &                                  {\footnotesize{(4.674)}} &                           -&       - &        {\footnotesize{(1.668)}} &             {\footnotesize{{(10.367)}}} &      {\footnotesize{(12.425)}} \\[1.0ex]
                                  
  & & - &         -0.011 &                0.079\threeS &       0.027\twoS &       0.108\threeS &                  0.169\threeS        \\
&     &               -             &    {\footnotesize{(-0.632)}}    &                         {\footnotesize{(4.988)}}&       {\footnotesize{(2.044)}} &  {\footnotesize{(10.415)}} &                   {\footnotesize{{(12.434)}}}  \\[1.0ex]

\cmidrule{3-8}
 &  \multirow{4}{*}{\text{t+66}}   &  0.027\threeS &                - &             - &        0.002 &             0.08\threeS &       0.217\threeS \\
 
     & &                                   {\footnotesize{(5.335)}} &             - &              -&       {\footnotesize{(0.259)}} &                     {\footnotesize{{(11.214)}}} &      {\footnotesize{(23.221)}} \\[1.0ex]
                                  
  & & - &         0.005 &                0.045\threeS &       0.006 &       0.087\threeS &                  0.216\threeS        \\
&     &               -             &    {\footnotesize{(0.475)}}    &                         {\footnotesize{(4.581)}}&       {\footnotesize{(0.703)}} &  {\footnotesize{(11.242)}} &                   {\footnotesize{{(23.242)}}}  \\[1.0ex]
\bottomrule
\multicolumn{3}{l}{\textsuperscript{***}$p<0.01$, 
  \textsuperscript{**}$p<0.05$, 
  \textsuperscript{*}$p<0.1$}
\end{tabular}                        
    \vspace{1em}
    \caption{Estimation results for HAR-RV and HAR-semiRV model specifications with daily leverage effect component. Results are displayed for the cryptocurrency and the equity panel over the four different horizons. T-stats are in parentheses. The model equation on top of the table reflects all the parameters common to the specification reported in this table. The total number of observations available considering all the entities used for the panel estimate is 47787 for crypto and 17111 for equity}
    \label{tab:HAR_leverage}
\end{table}
\end{singlespacing}
\subsection{The Role of Jumps}
The model specifications estimated in the previous subsections indicate that a positive realized variance has a greater impact on estimating future volatility than a negative realized variance. According to \cite{barndorff2008measuring}, this is due to differences in jump variation in a given time series. To account for this effect, we insert the signed jump variation measure, $\Delta J_t^2 = RV^{+}-RV^{-}$, to identify information from signed jumps. Such difference eliminates the integrated variance term and is positive when a day has a prevalence of upward jumps or negative when downward jumps are prevalent. \ref{tab:HAR_jumps} shows the estimation of the related model specification, including the jump variation term. The coefficient is always positive and statistically significant, except for the horizon $h=5$, for both asset classes. The sign of $\phi_j$ means that days dominated by negative jumps lead to lower future volatility, while days with positive jumps lead to higher future volatility. The jump variation estimated coefficients for the cryptocurrency cross-section are also larger than the equity cross-section, remarking on the more frequent presence of extreme market movements in the notoriously unstable and immature cryptocurrency market.

\ref{tab:HAR_jumps} also shows the results of the signed jump specification, which replaces the jump variation component with two signed components regarding the dominance of a positive or a negative jump, respectively. The coefficients for the positive jump variations are significant, except at $h=5$, and negative in sign, differently from the case of equity cross-section. Hence, this result says that a day dominated by a positive jump decreases future volatility, while a day dominated by a negative jump does the opposite. We conclude that the two signed jump components do not have the same effect. When isolated in two different factors, they provide additional information to interpret the effect of a jump. In this case, a negative jump is associated with an increase in future volatility, while a positive jump is with a decrease, only in the cryptocurrency case. Although this result seems to conflict with the inversion of the leverage effect in the cryptocurrency cross-section, the jump variation gives a perspective on the volatility movement at the intraday level. In contrast, the inverted leverage effect retrieved in the literature and the previous model specification accounts for a daily effect. These results highlight how the discrete component, i.e., the jump one, of the quadratic variation exhibits the asymmetric effect of return on future volatility, while, under the light of the previous section's results, the continuous component signals an inversion of such effect. The \ref{tab:HAR_NOjumps} in the appendix provides the estimated coefficients for model specifications that include only the continuous part of the volatility for the most recent lags, without any jumps.

The presence of a leverage effect guaranteed by the discrete part of the quadratic variation points to different categories of operators in the cryptocurrency market. As outlined in the previous section, the inverse leverage effect is caused by uninformed investors that permeate the cryptocurrency market. Indeed, this is a market that, unlike more traditional markets, has seen a huge exposure to retail investors, while the institutional and professional operators came after. Hence, we explain the different behaviors of the volatility drivers by associating them with a specific category of market operators. The activity of high-frequency professional traders regulates the volatility dynamics according to the renowned leverage effect because the price variation associated with jumps is transient, so only they can exploit such information to make an arbitrage. On the other hand, some less-informed retail investors operate under the FoMO guide and inverse the leverage direction.

\begin{singlespacing}
\begin{table}[H]
    \caption*{\small{$RV_{h,t+h} = \mu + \phi_j \Delta J^2_t +   \phi_j- \Delta J^{2-}_t + \phi_j+ \Delta J^{2+}_t + \phi_c BV_t + \phi_w RV_{w,t} + \phi_m RV_{m,t} + \epsilon_{t+h}$}}
\centering
\begin{tabular}{cccccccc} 
\toprule
        &       & $\mathbf{\phi_j}$ &  $\mathbf{\phi_j^-}$ &  $\mathbf{\phi_j^+}$  &  $\mathbf{\phi_c}$  & $\mathbf{\phi_w}$ & $\mathbf{\phi_m}$  \\
\midrule
 \multirow{16}{*}{\rotatebox[origin=c]{90}{Crypto}} & \multirow{4}{*}{\text{t+1}} &       0.554\threeS &                - &       - &  0.668\threeS &                        0.008\oneS &       0.045\threeS \\
 
     &                            &       {\footnotesize{(6.867)}} &             - &       - & {\footnotesize{(12.469)}} &                    {\footnotesize{{(1.783)}}} &      {\footnotesize{(5.798)}} \\[1.0ex]
                                 
&  &         - &               0.643\threeS &       -0.13\twoS &       0.341\threeS &                  0.011\threeS &       0.039\threeS \\
  
     &                            &    -    &                         {\footnotesize{(4.565)}}&       {\footnotesize{(-2.298)}} &  {\footnotesize{(3.593)}}  &                   {\footnotesize{{(3.061)}}} &      {\footnotesize{(4.149)}} \\[1.0ex]

\cmidrule{3-8}
& \multirow{4}{*}{\text{t+5}} &       -0.039 &                - &            - &       0.359\threeS &             0.042\oneS &       0.025 \\
 
     &                           &       {\footnotesize{(-0.085)}} &                           -&       - &      {\footnotesize{(3.809)}}  &              {\footnotesize{{(1.982)}}} &      {\footnotesize{(0.964)}} \\[1.0ex]
                                
 & &         - &               -0.022 &       -0.083 &             0.186 &            0.051\twoS &       0.024 \\
  
      &                           &    -    &                         {\footnotesize{(-0.037)}}&       {\footnotesize{(-0.93)}} &   {\footnotesize{(1.187)}}  &                  {\footnotesize{{(2.304)}}} &      {\footnotesize{(0.843)}} \\[1.0ex]

\cmidrule{3-8}
 &\multirow{4}{*}{\text{t+22}} &       0.144\twoS &                - &             - &          0.099\threeS &           0.026\twoS &       0.042\threeS \\
 
     &                             &       {\footnotesize{(2.294)}} &                            -&       - &           {\footnotesize{(6.684)}} &          {\footnotesize{{(2.495)}}} &      {\footnotesize{(3.697)}} \\[1.0ex]
                                  
 & &         - &                0.243\twoS &      -0.044\oneS &           0.112\threeS &              0.025\twoS &       0.039\threeS \\
  
    &                             &    -    &                         {\footnotesize{(2.533)}}&       {\footnotesize{(-1.828)}} & {\footnotesize{(4.408)}}  &                    {\footnotesize{{(2.41)}}} &      {\footnotesize{(3.461)}} \\[1.0ex]

\cmidrule{3-8}
 &\multirow{4}{*}{\text{t+66}} &       0.107\threeS &                - &             - &        0.044\threeS &             0.005 &       -0.016\threeS \\
 
    &                              &       {\footnotesize{(7.78)}} &              - &             -&       {\footnotesize{(10.685)}} &                     {\footnotesize{{(1.072)}}} &      {\footnotesize{(-3.994)}} \\[1.0ex]
                                  
 & &         - &                0.19\threeS &       -0.029\twoS &            0.07\threeS &            0.004 &       -0.017\threeS \\
  
     &                            &    -    &                         {\footnotesize{(9.454)}}&       {\footnotesize{(-2.53)}} &             {\footnotesize{(9.073)}}  &        {\footnotesize{{(0.939)}}} &      {\footnotesize{(-4.407)}} \\[1.0ex]
                                  
\midrule
  \multirow{16}{*}{\rotatebox[origin=c]{90}{Equity}} & \multirow{4}{*}{\text{t+1}} &   0.061\twoS &          - &      - &       1.076\threeS &                          0.129\threeS &       0.129\threeS \\
 
     &   &                               {\footnotesize{(2.384)}} &             - &       - &           {\footnotesize{(9.221)}} &          {\footnotesize{{(5.398)}}} &      {\footnotesize{(2.602)}} \\[1.0ex]
                                 
 &   &       - &                0.003 &       0.154\twoS &        1.002\threeS &                 0.132\threeS &       0.131\threeS \\
  
  &    &                              -    &                         {\footnotesize{(0.099)}}&       {\footnotesize{(2.197)}} &              {\footnotesize{(8.061)}}  &       {\footnotesize{{(5.472)}}} &      {\footnotesize{(2.661)}} \\[1.0ex]

\cmidrule{3-8}
   &   \multirow{4}{*}{\text{t+5}}   & 0.017 &                - &            - &       0.575\threeS &             0.148\threeS &       0.142\threeS \\
 
   &  &                                  {\footnotesize{(0.937)}} &                           -&       - &        {\footnotesize{(11.106)}} &             {\footnotesize{{(10.182)}}} &      {\footnotesize{(5.53)}} \\[1.0ex]
                                
    & &      - &                0.055 &       0.07\twoS &             0.552\threeS &            0.148\threeS &       0.14\threeS \\
  
     &  &                              -    &                         {\footnotesize{(0.24)}}&       {\footnotesize{(2.212)}} &              {\footnotesize{(10.33)}}  &       {\footnotesize{{(10.174)}}} &      {\footnotesize{(5.476)}} \\[1.0ex]

\cmidrule{3-8}
  & \multirow{4}{*}{\text{t+22}}   &  0.035\threeS &                - &             - &        0.335\threeS &            0.088\threeS &       0.154\threeS \\
 
     &  &                                  {\footnotesize{(8.769)}} &                 - &           -&      {\footnotesize{(9.745)}} &                     {\footnotesize{{(8.664)}}} &      {\footnotesize{(11.539)}} \\[1.0ex]
                                  
     & &     - &                0.021\threeS &      0.067\threeS &            0.307\threeS &             0.088\threeS &       0.154\threeS \\
  
    &  &                               -    &                         {\footnotesize{(2.279)}}&       {\footnotesize{(2.791)}} &          {\footnotesize{(8.634)}}  &           {\footnotesize{{(8.722)}}} &      {\footnotesize{(11.553)}} \\[1.0ex]

\cmidrule{3-8}
 &  \multirow{4}{*}{\text{t+66}}   &  0.021\threeS &                - &             - &        0.21\threeS &             0.067\threeS &       0.205\threeS \\
 
     & &                                   {\footnotesize{(3.627)}} &             - &              -&       {\footnotesize{(9.01)}} &                     {\footnotesize{{(9.55)}}} &      {\footnotesize{(22.214)}} \\[1.0ex]
                                  
& & - &                0.018\threeS &      0.028\twoS &             0.205\threeS &            0.067\threeS &       0.205\threeS \\
  
    &      &                            -    &                         {\footnotesize{(2.744)}}&       {\footnotesize{(2.093)}} &            {\footnotesize{(8.528)}}  &         {\footnotesize{{(9.563)}}} &      {\footnotesize{(22.213)}} \\[1.0ex]
\bottomrule
\multicolumn{3}{l}{\textsuperscript{***}$p<0.01$, 
  \textsuperscript{**}$p<0.05$, 
  \textsuperscript{*}$p<0.1$}
\end{tabular}                        
    \vspace{1em}
    \caption{Estimation results for HAR-JV and HAR-SJV model specifications. Results are displayed for the cryptocurrency and the equity panel over the four different horizons. T-stats are in parentheses. The model equation on top of the table reflects all the parameters common to the specification reported in this table. The total number of observations available considering all the entities used for the panel estimate is 47787 for crypto and 17111 for equity}
    \label{tab:HAR_jumps}
\end{table}
\end{singlespacing}

\subsection{Behavior under different market trends}
\begin{figure}
 \centering\includegraphics{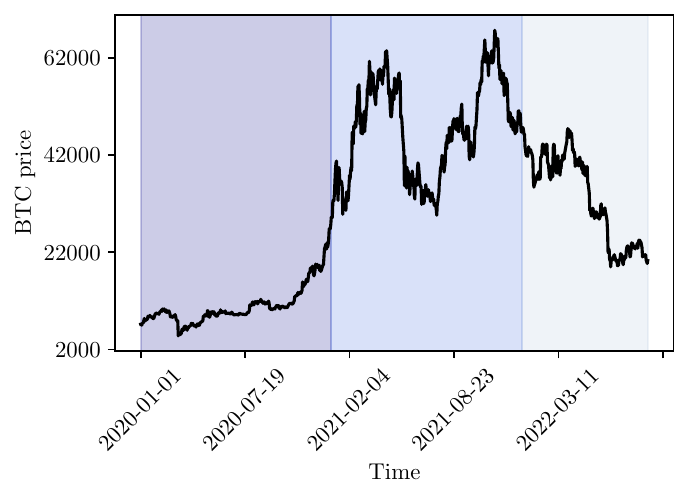}
 \vspace*{-5mm}
 \caption{BTC price dynamics over the historical period under analyses. The shades of blue separate the time series year by year, highlighting the different market regimes in the cryptocurrency market.}
 \label{Fig:btc_dynamics}
\end{figure}

The cryptocurrency market underwent several changes in the regime due to the developing nature of this ecosystem and the underlying global economic condition. This aspect is especially true when considering the market run from 2020 to mid-year 2021 when many new protocols had been released, and terms like DeFi protocol or Non-fungible Token (NFT) have been popularized among a larger audience. Using BTC, the largest capitalized cryptocurrency, as a proxy for the whole market dynamics, one can observe from  \ref{Fig:btc_dynamics} different market trends: a sharp increase in 2020, two major drawdowns in 2021, and a steep decrease in 2022. We aim to estimate the same volatility model specifications of the previous sections under those different market regimes to validate the robustness of the previous sections' findings. Hence, in this Section, we provide the estimates considering just the windows 2020-2021 and 2021-2022 and the whole period covered by the data, 2020-2022, which is available in the previous Sections. The results of the estimated coefficients provided in this Section are available in the table format in the appendix, where statistical significance can be assessed by looking at the resulting t-statistics. Considering all the entities used for the panel estimate, the total number of observations available is 29005 for crypto and 11472 for equity in the window 2020-2021 and 45191 for crypto and 15968 for equity in the window 2021-2022. The number of observations for the whole dataset (2020-2022) is the same as provided in the previous subsection.

 \FloatBarrier

\begin{figure}
\centering
\begin{subfigure}{.5\textwidth}
  \centering
  \includegraphics[width=\linewidth]{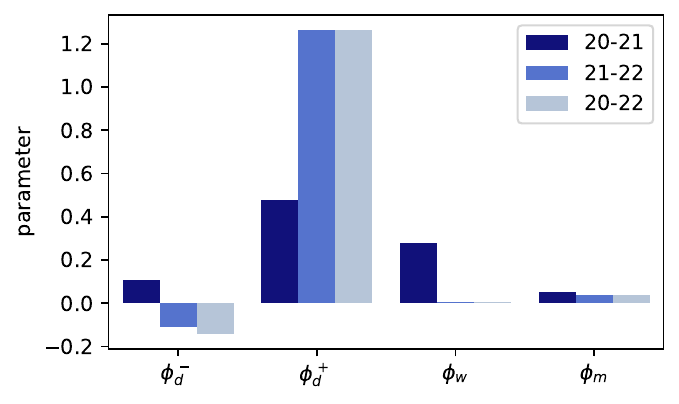}
\end{subfigure}%
\begin{subfigure}{.5\textwidth}
  \centering
  \includegraphics[width=\linewidth]{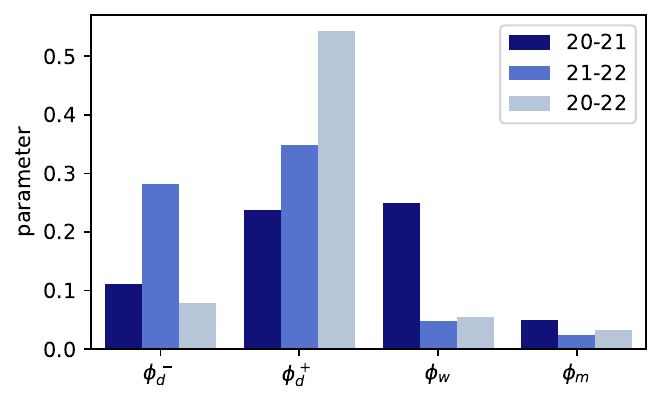}
\end{subfigure}
\begin{subfigure}{.5\textwidth}
  \centering
  \includegraphics[width=\linewidth]{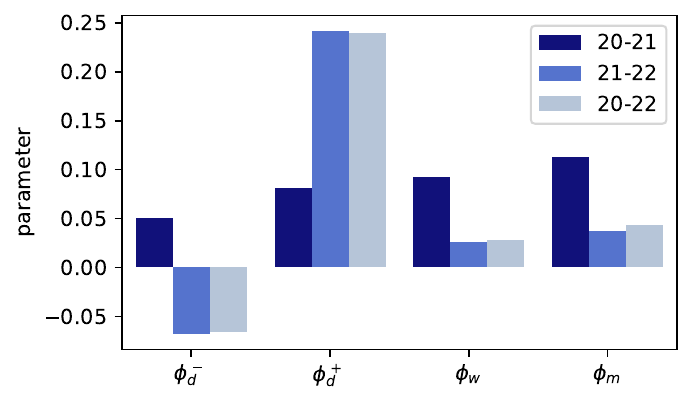}
\end{subfigure}%
\begin{subfigure}{.5\textwidth}
  \centering
  \includegraphics[width=\linewidth]{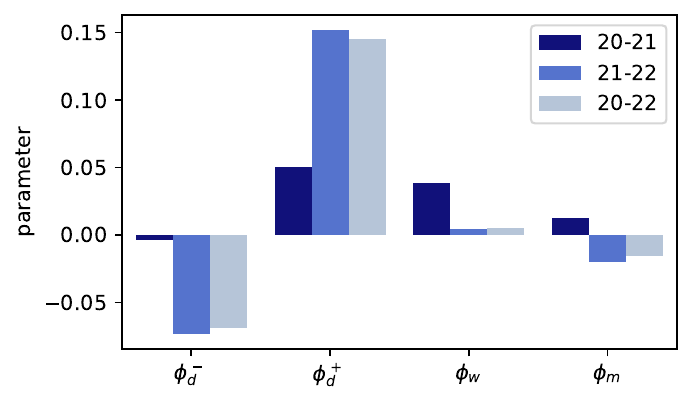}
\end{subfigure}
 \caption{Barplots of estimated coefficient for the panel HAR-semiRV model specification over three different time windows. The number in the legend reflects the year of analysis and is inclusive. Each subplot differs for the future horizon of the model specifications: 1 day (top left), 5 days (top right), 22 days (bottom left), and 66 (bottom right).}
  \label{Fig:DecRV1_trends}
\end{figure}

The Figures of this Section comprise four subplots, one for each horizon. The one-day ahead (first plot) and the five-day ahead (second plot) horizons are in the first row, while the one-month ahead (first plot) and the three-month ahead (second plot) are in the second row. \ref{Fig:DecRV1_trends} shows the estimation for the panel HAR model when the first regressor is decomposed into its signed components. The first remarkable effect relies on the volatility's more considerable persistence in the cryptocurrency market for the two years 2020-2021, where the weekly component of the model plays a role in the volatility estimation at any of the time horizons tested. The monthly components in that period are less significant in magnitude. Still, it is significant and more prominent than the other estimation windows, where the estimated coefficient $\phi_m$ is more marginal. In addition, by looking at the parameter $\phi^+_d$, we observe that the impact of the previous-day positive volatility is less pronounced with respect to the whole estimation period or the subsequent window that spans 2021-2022. The reason for this effect can be retrieved in the extremely bullish market regime (see \ref{Fig:btc_dynamics}) when no sudden market drawdown occurred. That two-year period experienced a majority of positive returns, making it more difficult to visualize the effect compared to the adjacent period we have considered in the analyses. Given the similarity of the estimation results for those two estimation windows, when compared to the 2020-2021 window, we conclude that a large amount of the effect of the positive part of the recent volatility is due to the most recent period in 2022. This result indicates that the bear market regime in 2022 signaled market activity in response to upward movements at the intraday level. Recalling the results from the previous subsection, this effect is similar to the equity cross-section but larger in magnitude. Such results are opposite to what was retrieved from results in previous works that employed the realized semivariances as volatility proxies \citep{fang2017realized,bollerslev2020multivariate}, where the negative part of the intraday volatility provides a relevant contribution to future volatility. Under this view, cryptocurrencies show different traits in the volatility pattern compared to other, more established asset classes due to a clear separation of behaviors among informed traders and uninformed retail investors that participate in the market. 

\FloatBarrier

\begin{figure}
\centering
\begin{subfigure}{.5\textwidth}
  \centering
  \includegraphics[width=\linewidth]{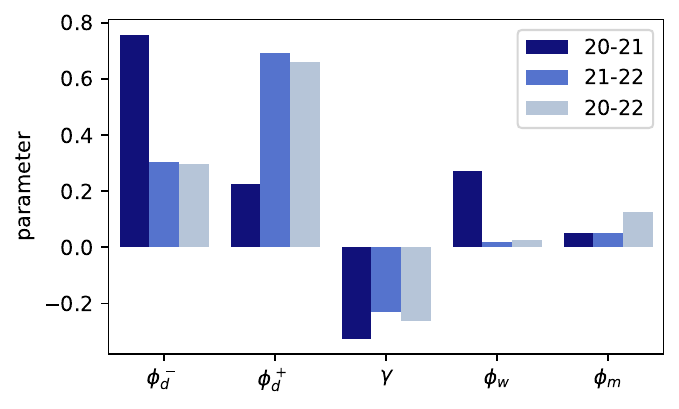}
\end{subfigure}%
\begin{subfigure}{.5\textwidth}
  \centering
  \includegraphics[width=\linewidth]{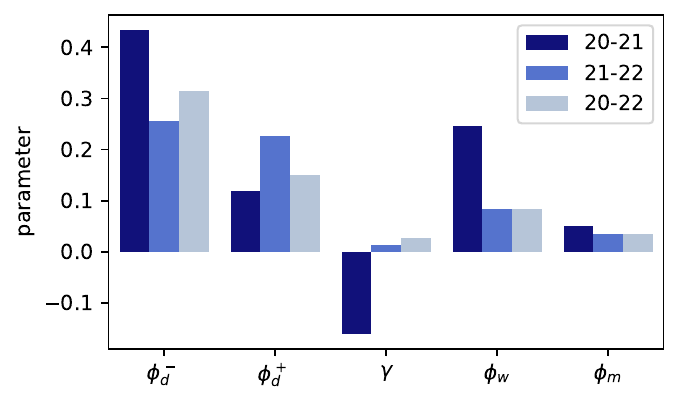}
\end{subfigure}
\begin{subfigure}{.5\textwidth}
  \centering
  \includegraphics[width=\linewidth]{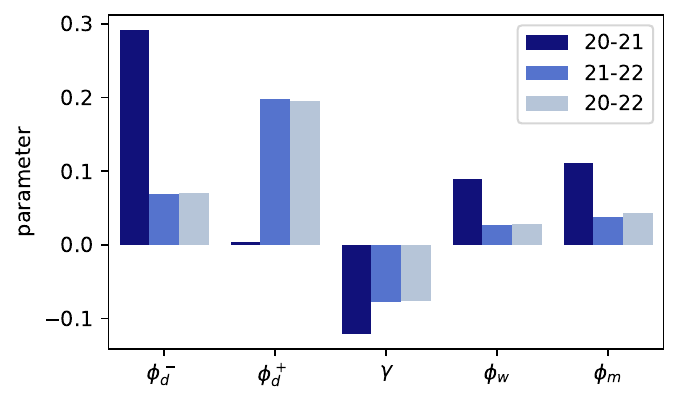}
\end{subfigure}%
\begin{subfigure}{.5\textwidth}
  \centering
  \includegraphics[width=\linewidth]{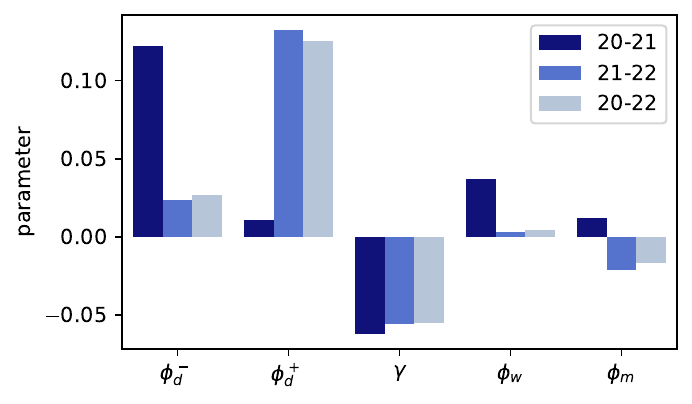}
\end{subfigure}
 \caption{Barplots of estimated coefficients for the panel HAR-semiRV model specification with daily leverage effect over three different time windows. The number in the legend reflects the year of analysis and is inclusive. Each subplot differs for the future horizon of the model specifications: 1 day (top left), 5 days (top right), 22 days (bottom left), and 66 (bottom right).}
  \label{Fig:DecRV1NegRet_trends}
\end{figure}

Moving from the intraday effect to the daily leverage effect, \ref{Fig:DecRV1NegRet_trends} shows the decomposed first lag specification estimation results when the latter effect is added. The $\gamma$ parameter is consistently significant and with the same magnitude for all the estimation windows tested, except for the horizon equal to five days ahead. The coefficient is always negative, reflecting the inversion of the leverage effect retrieved daily. As previously studied in \cite{baur2018asymmetric,brini2022assessing}, negative returns' impact tends to lower the cryptocurrency volatility instead of increasing it, as is commonly expected from the empirical literature. The effect is robust, even though more pronounced during the overall bull market of the 2020-2021 period. Also, in this model specification, the volatility appears more persistent in the two years, 2020-2021, than in the other estimation period.

\FloatBarrier

\begin{figure}
\centering
\begin{subfigure}{.5\textwidth}
  \centering
  \includegraphics[width=\linewidth]{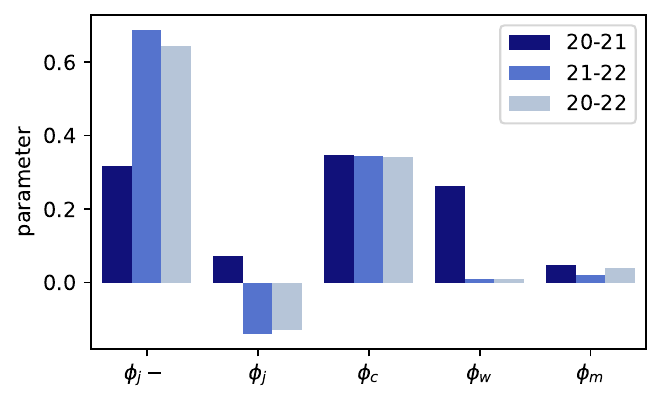}
\end{subfigure}%
\begin{subfigure}{.5\textwidth}
  \centering
  \includegraphics[width=\linewidth]{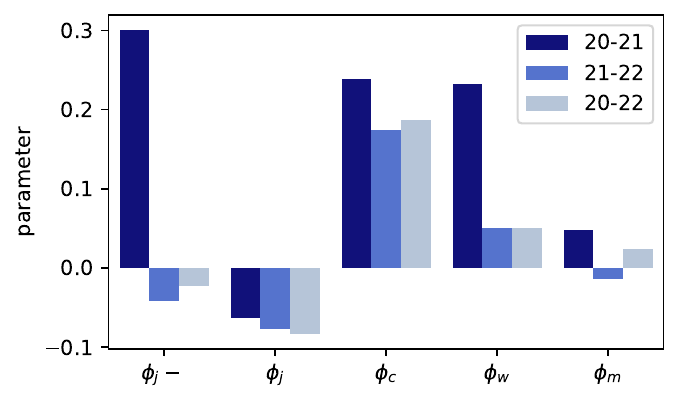}
\end{subfigure}
\begin{subfigure}{.5\textwidth}
  \centering
  \includegraphics[width=\linewidth]{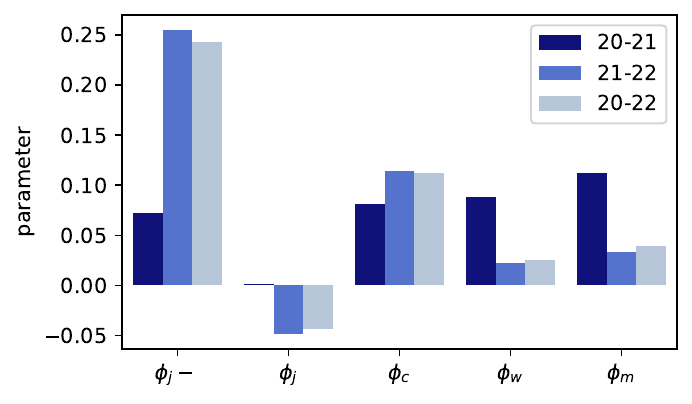}
\end{subfigure}%
\begin{subfigure}{.5\textwidth}
  \centering
  \includegraphics[width=\linewidth]{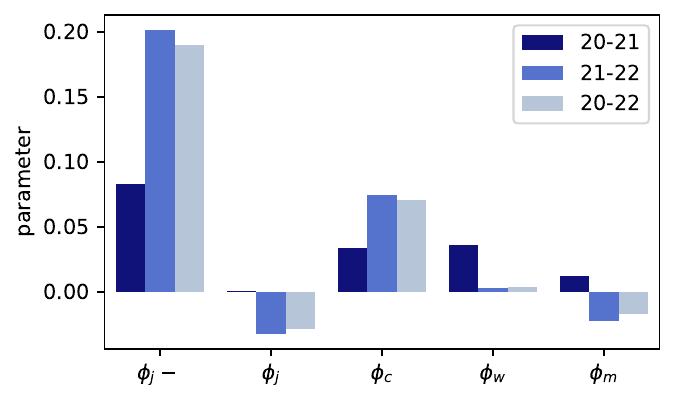}
\end{subfigure}
 \caption{Barplots of estimated coefficient for the panel HAR-JV and panel HAR-SJV model specification over three different time windows. The number in the legend reflects the year of analysis and is inclusive. Each subplot differs for the future horizon of the model specifications: 1 day (top left), 5 days (top right), 22 days (bottom left), and 66 (bottom right).}
  \label{Fig:decSJV_trends}
\end{figure}

\ref{Fig:decSJV_trends} provided the result of the role of the most recent signed jump components over the different estimation windows. As for the whole signed volatility components, the jump one presents a less influential impact of the $\phi_j^-$ when estimated over 2020-2021, except for the horizon equal to five. Negative jumps happen to be more influential due to the bear market of 2022, while they are more negligible during the upward trend of 2020. Also, the signed jump specification shows a different level of volatility persistence among the selected estimation windows. In the appendix, \ref{Fig:DecRV_trends} and \ref{Fig:BVnojump_trends} show the estimated coefficients for a fully decomposed HAR-RV model into its signed components and a HAR-SJV that includes only the continuous part of the volatility for the most recent lags, respectively.

\FloatBarrier

The results of this section show the cryptocurrency ecosystem's evolution from a quantitative perspective. The structural change in the cryptocurrency ecosystem from the beginning of 2020 is one reason for the decrease in volatility persistence. The proliferation of new coins and tokens, the emergence of different blockchains to challenge the most established Ethereum blockchain, in terms of fees and validation speed, and the entrance of institutional investors in the market contributed significantly to the observed ecosystem growth \citep{bialkowski2020cryptocurrencies}. Such an effect is particularly evident in the traded volume and the number of active participants in the space \citep{bouri2019trading,bains2022regulating}. 
\subsection{Disentangling the effects at the individual level}\label{Sec:individual}
After analyzing the cryptocurrency ecosystem at the macro level, we analyze the data in our universe at the micro level by individually estimating each model's specifications for each entity. Since the panel estimation provides an average effect of the whole ecosystem, we are also interested in evaluating the tested effects at the individual level to understand which results in the previous Sections are commonly shared among many universe components. Similar to the previous section, each figure presents a group of subplots, each referring to a different estimation horizon and following the same order. We present the signed volatility component results, the daily leverage effect, and the signed jump components. For each bar plot, the cryptocurrencies are ranked by the magnitude of the respective coefficient, and the bars are either blue if the estimation is statistically significant at least at 5\%, or red if they are not. For some cryptocurrencies, the individual estimation of the HAR-like model does not converge; hence, these cryptocurrencies are excluded from the graphic representation.

We acknowledge the challenges and limitations of applying OLS estimation to individual time series within the panel. In our methodological approach, we recognize that the properties of the error term can differ between panel and time series data. In panel data, the error term may have a complex structure, including both individual-specific and time-specific effects. This complexity can sometimes be averaged out in panel estimations but might cause problems in individual time series estimations. Moreover, when dealing with individual time series, especially with short-time dimensions, OLS can be biased due to the presence of lagged dependent variables as regressors, of which the HAR-like specifications are an example. This bias might not be as pronounced when considering a panel estimation, and it is one of the reasons for not having all the individual estimations converge. As a third motivation for the OLS convergence problem of certain individual model specifications, we note that in a panel dataset, the OLS method benefits from the larger sample size, as it combines cross-sectional and time series data. This increased sample size can lead to more stable and reliable estimates. In contrast, individual time series may have fewer observations, leading to less stable estimates and potential convergence issues. Under this view, panel data often captures heterogeneous units (different cryptocurrencies in our case) over time. When you estimate the model for the entire panel, the OLS method captures the average effect across all units. However, when you analyze individual time series, the specific characteristics of each unit come into play. If a particular time series has unique features that are not well-captured by the OLS assumptions (like non-linearity, or structural breaks), the estimation might fail to converge. Moreover, it allows us to identify any outliers or influential observations that can drive the panel regression results and contribute to a more granular understanding of the volatility pattern for each cryptocurrency considered rather than treating them as a homogeneous group.

\begin{figure}[h]
    \centering
    \includegraphics[width=0.8\textwidth]{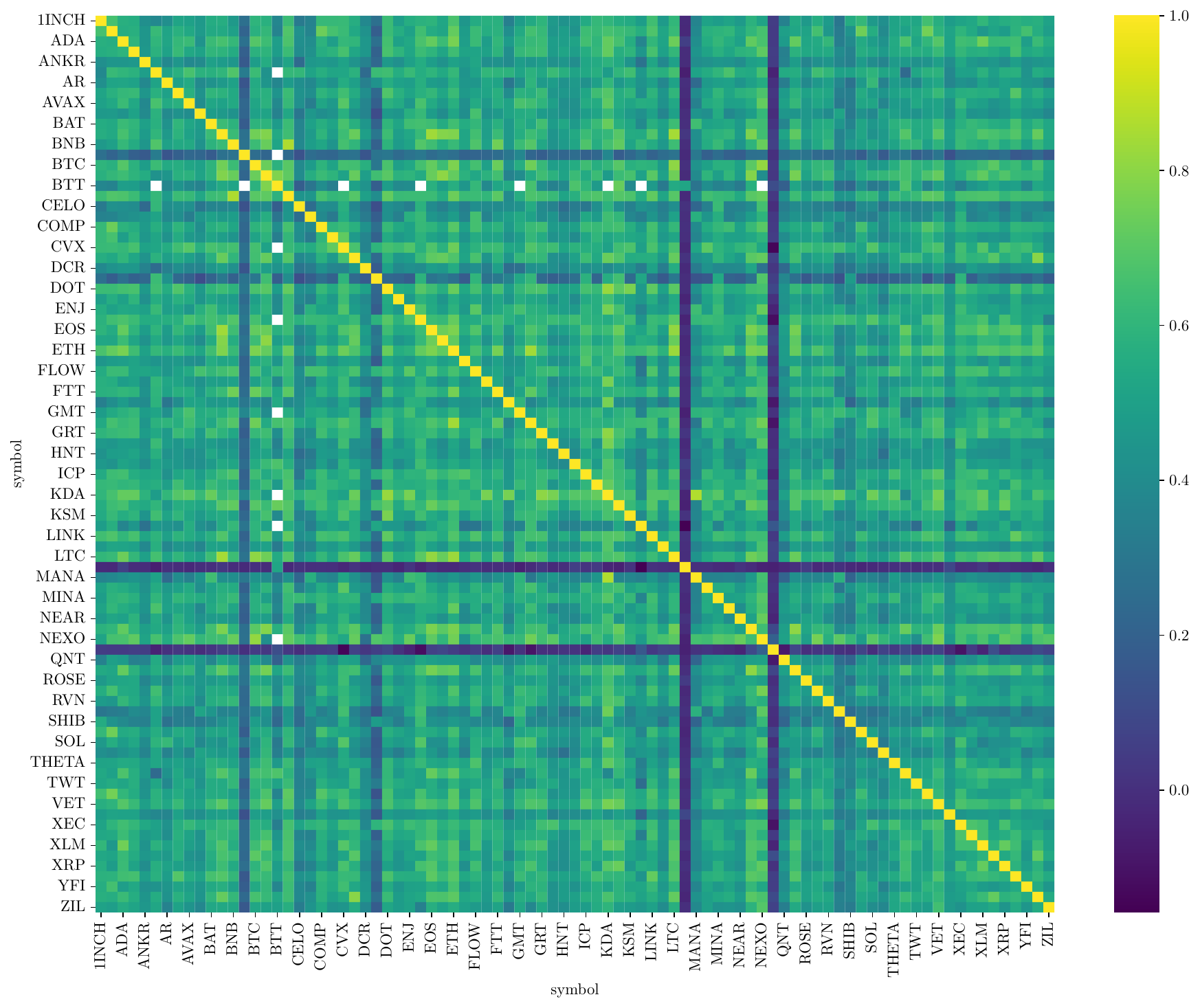}
    \caption{Cryptocurrency daily return correlation over the 2020-2022 period.}
    \label{Fig:retcorr}
\end{figure}

Despite this, the individual estimation procedure can provide additional insights into the heterogeneity of the entities and the robustness of the panel regression results. The cryptocurrency market is distinguished by its pronounced volatility but also by the heterogeneity of its constituents. Some of them are established cryptocurrencies such as BTC and ETH, and others have only relatively recent history as they did their Initial Coin Offering (ICO) \citep{ante2018blockchain,ashta2018fintech} or started being traded on Binance (the data source) shortly before or after 2020. The idea of heterogeneity is often overlooked when one thinks about the high level of correlation among cryptocurrencies since the most capitalized one usually drags the market dynamics, resulting in a correlation heatmap for the daily returns over the period analyzed as in \ref{Fig:retcorr}.

\FloatBarrier

However, each cryptocurrency is usually backed by a different project idea or use case, which makes it interesting to disentangle the aggregated effect provided by the pooled panel analysis. Our decision to undertake individual estimations for each cryptocurrency is driven by the intention to investigate the distinct nature, if any, of the cryptocurrency market, characterized by its diversity and the unique backing of each currency by different project ideas or use cases. Although we recognize the limitations inherent in using price data alone, where the assumption that prices immediately reflect underlying economic, technological, and regulatory developments may not hold due to the known latency in market reactions, our decision to perform individual estimations for each cryptocurrency is primarily driven by the objective to investigate the distinct price patterns. This methodological choice allows us to explore the nuances of price movements and volatility patterns across different cryptocurrencies, acknowledging that these price dynamics may serve as proxies for underlying determinants in the absence of more detailed data\footnote{There is no established consensus for one or the other approach, and they can be used concurrently depending on the use case \citep{bou2018univariate}. This comparison has been further analyzed within the literature at the intersection of statistics and social sciences \citep{allison1997change,teachman2001covariance,ejrnaes2006comparing,allison2009fixed}.}.

\begin{figure}
\centering
\begin{subfigure}{.5\textwidth}
  \centering
  \includegraphics[width=\linewidth]{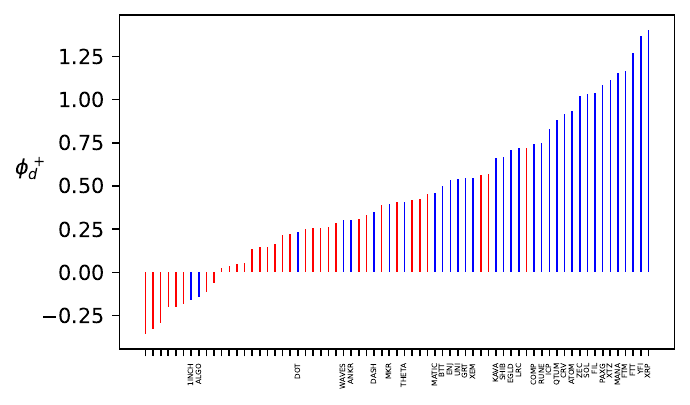}
\end{subfigure}%
\begin{subfigure}{.5\textwidth}
  \centering
  \includegraphics[width=\linewidth]{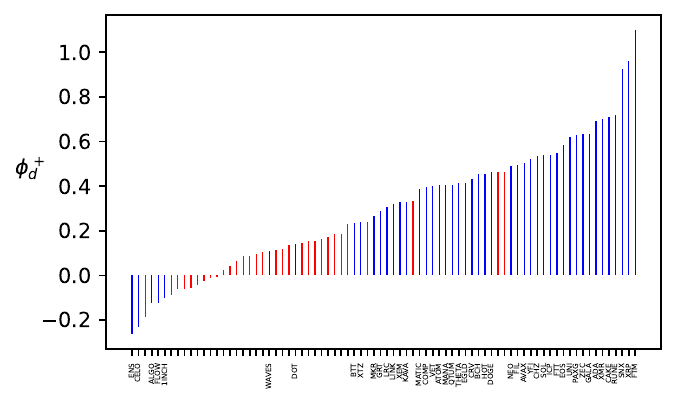}
\end{subfigure}
\begin{subfigure}{.5\textwidth}
  \centering
  \includegraphics[width=\linewidth]{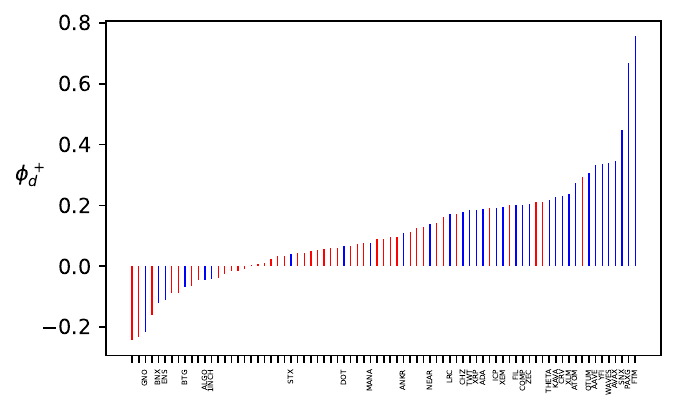}
\end{subfigure}%
\begin{subfigure}{.5\textwidth}
  \centering
  \includegraphics[width=\linewidth]{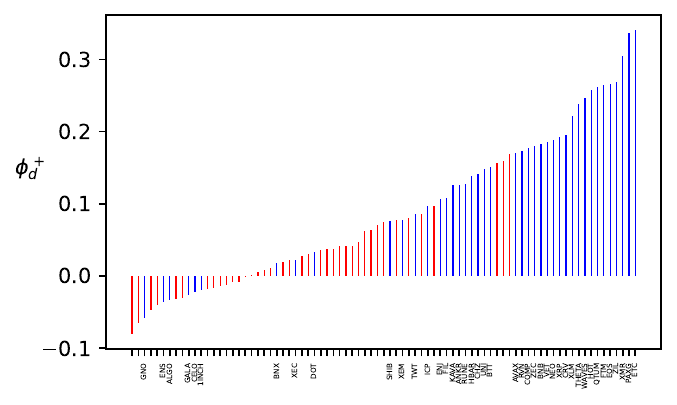}
\end{subfigure}
 \caption{Barplots of $\phi^+_d$ coefficients for the individual HAR-semiRV model specification over three different time windows. Each subplot differs for the future horizon of the model specifications: 1 day (top left), 5 days (top right), 22 days (bottom left), and 66 (bottom right). Blue bars identify statistically significant parameters at 5\% confidence level, while red bars are for statistically insignificant ones.}
  \label{Fig:indiv_DecRV1_pos}
\end{figure}

\begin{figure}
\centering
\begin{subfigure}{.5\textwidth}
  \centering
  \includegraphics[width=\linewidth]{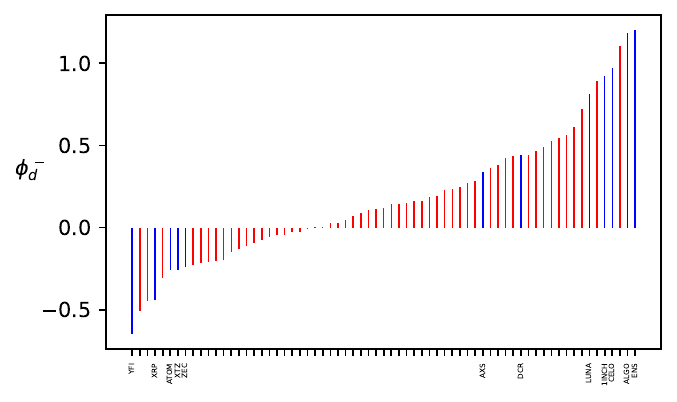}
\end{subfigure}%
\begin{subfigure}{.5\textwidth}
  \centering
  \includegraphics[width=\linewidth]{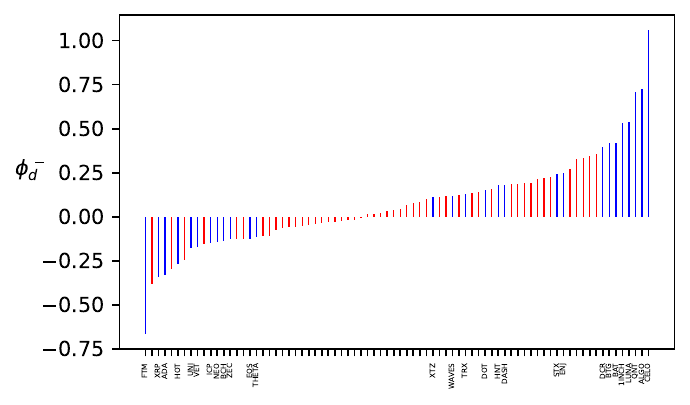}
\end{subfigure}
\begin{subfigure}{.5\textwidth}
  \centering
  \includegraphics[width=\linewidth]{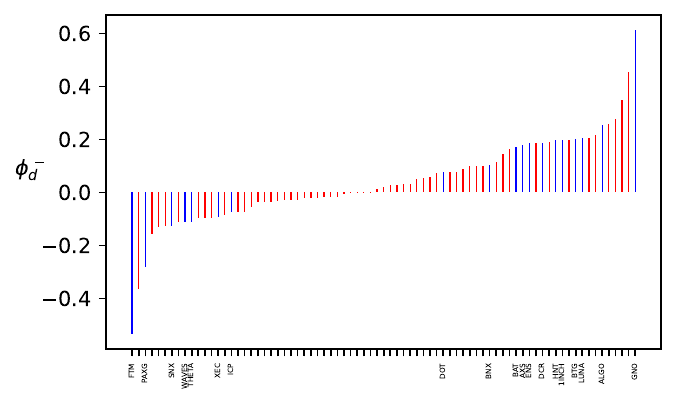}
\end{subfigure}%
\begin{subfigure}{.5\textwidth}
  \centering
  \includegraphics[width=\linewidth]{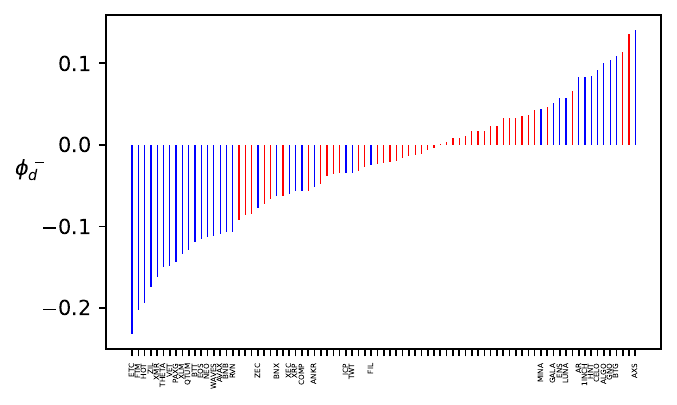}
\end{subfigure}
 \caption{Barplots of $\phi^-_d$ coefficients for the individual HAR-semiRV model specification over three different time windows. Each subplot differs for the future horizon of the model specifications: 1 day (top left), 5 days (top right), 22 days (bottom left), and 66 (bottom right). Blue bars identify statistically significant parameters at 5\% confidence level, while red bars are for statistically insignificant ones.}
  \label{Fig:indiv_DecRV1_neg}
\end{figure}

\ref{Fig:indiv_DecRV1_pos} and \ref{Fig:indiv_DecRV1_neg} show the individual estimates for the realized semivariances parameters, respectively, the positive and negative part at any of the previously mentioned future horizons. The significant estimated parameters $\phi^+_d$ have the same sign for most of the sample under study, with few exceptions at any horizon with negative signs. On the contrary, the significant estimated parameters $\phi^-_d$ are less in number, and the individual results are contrasting, with some of the cryptocurrencies in the universe having a positive impact for the negative realized semivariance and some others having a negative one. Those results validate the estimation of the panel models in \ref{Sec:signret} where the effect of $\phi^+_d$ impacts more than $\phi^-_d$ on future volatilities.

\FloatBarrier

\FloatBarrier

\begin{figure}
\centering
\begin{subfigure}{.5\textwidth}
  \centering
  \includegraphics[width=\linewidth]{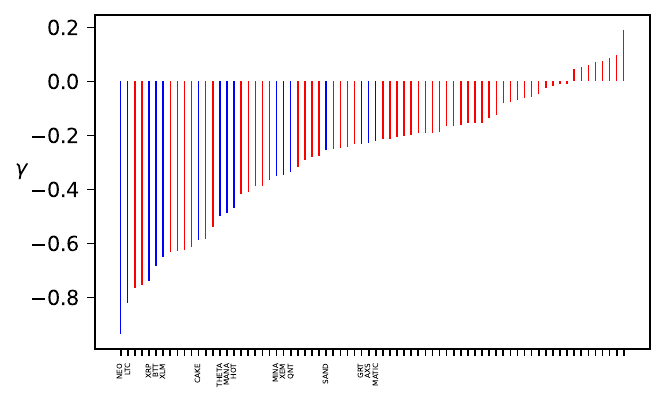}
\end{subfigure}%
\begin{subfigure}{.5\textwidth}
  \centering
  \includegraphics[width=\linewidth]{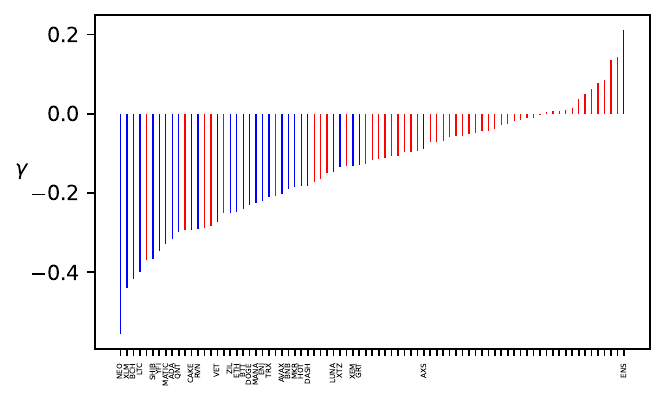}
\end{subfigure}
\begin{subfigure}{.5\textwidth}
  \centering
  \includegraphics[width=\linewidth]{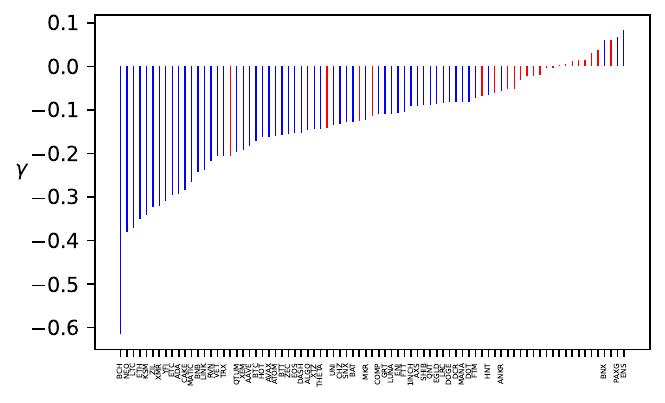}
\end{subfigure}%
\begin{subfigure}{.5\textwidth}
  \centering
  \includegraphics[width=\linewidth]{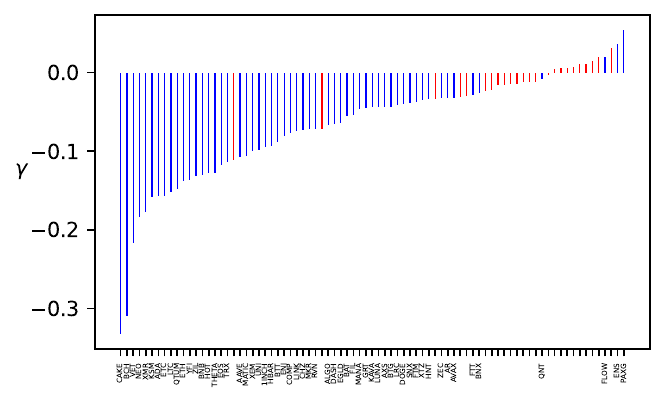}
\end{subfigure}
 \caption{Barplots of $\gamma$ coefficients for the individual HAR-semiRV model specification with daily leverage effect over three different time windows. Each subplot differs for the future horizon of the model specifications: 1 day (top left), 5 days (top right), 22 days (bottom left), and 66 (bottom right). Blue bars identify statistically significant parameters at 5\% confidence level, while red bars are for statistically insignificant ones.}
  \label{Fig:indiv_NegRet}
\end{figure}

Differently from $\phi^+_d$ and $\phi^-_d$, where the percentage of significant estimates in the selected cryptocurrency universe does not change with the model horizons, the daily leverage effect parameter $\gamma$ tends to be significant for the majority of the entities when the time horizon is larger, one-month or three-month ahead. \ref{Fig:indiv_NegRet} showcases these results, highlighting that the significant parameters $\gamma$ are always negative with few exceptions.

\FloatBarrier

\begin{figure}
\centering
\begin{subfigure}{.5\textwidth}
  \centering
  \includegraphics[width=\linewidth]{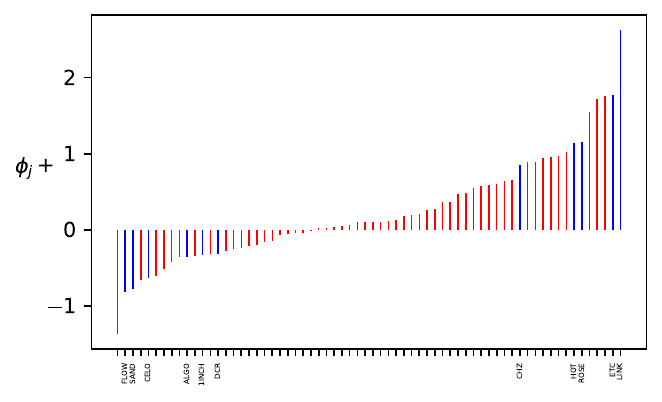}
\end{subfigure}%
\begin{subfigure}{.5\textwidth}
  \centering
  \includegraphics[width=\linewidth]{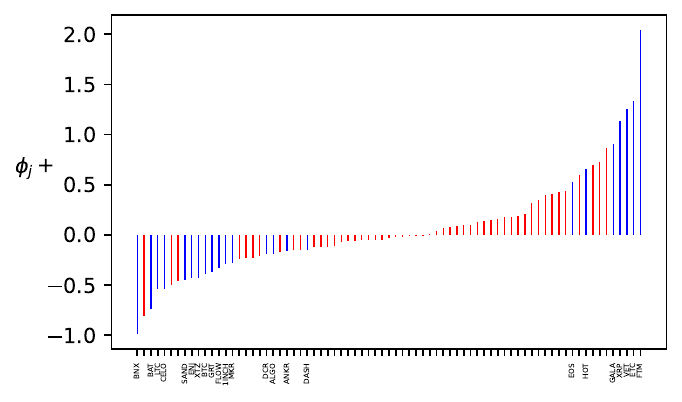}
\end{subfigure}
\begin{subfigure}{.5\textwidth}
  \centering
  \includegraphics[width=\linewidth]{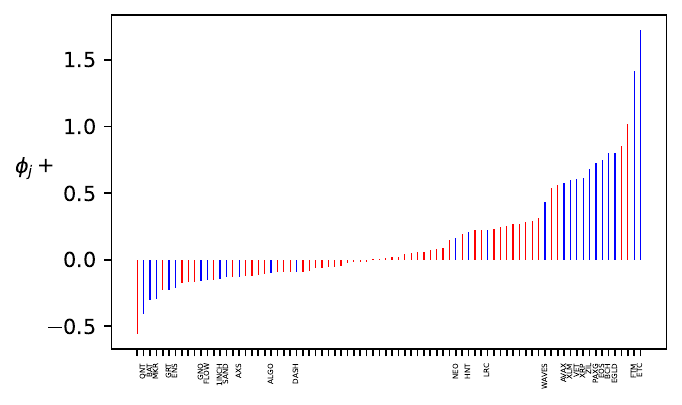}
\end{subfigure}%
\begin{subfigure}{.5\textwidth}
  \centering
  \includegraphics[width=\linewidth]{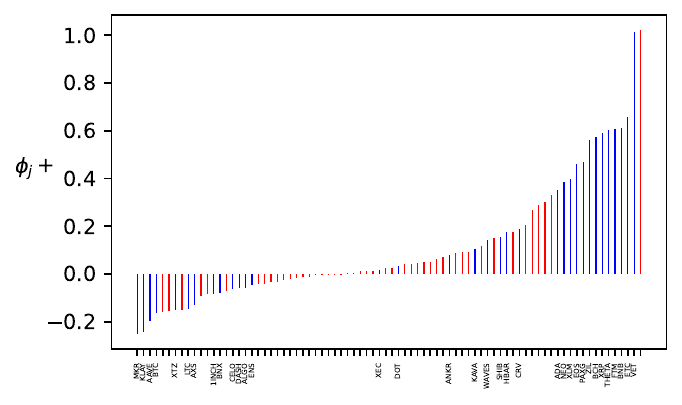}
\end{subfigure}
 \caption{Barplots of $\phi_j^+$ coefficients for the individual HAR-SJV model specification over three different time windows. Each subplot differs for the future horizon of the model specifications: 1 day (top left), 5 days (top right), 22 days (bottom left), and 66 (bottom right). Blue bars identify statistically significant parameters at 5\% confidence level, while red bars are for statistically insignificant ones.}
  \label{Fig:indiv_decSJV_pos}
\end{figure}

\begin{figure}
\centering
\begin{subfigure}{.5\textwidth}
  \centering
  \includegraphics[width=\linewidth]{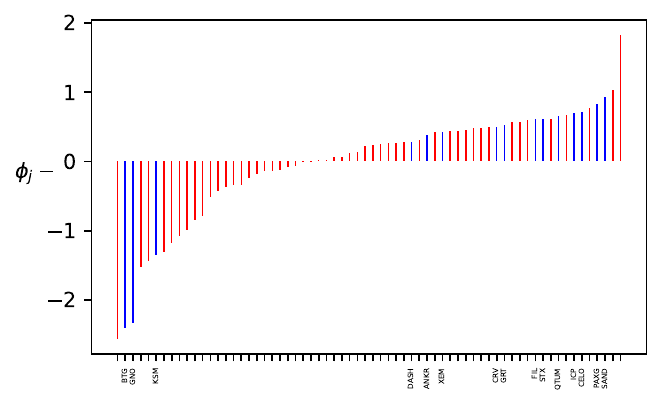}
\end{subfigure}%
\begin{subfigure}{.5\textwidth}
  \centering
  \includegraphics[width=\linewidth]{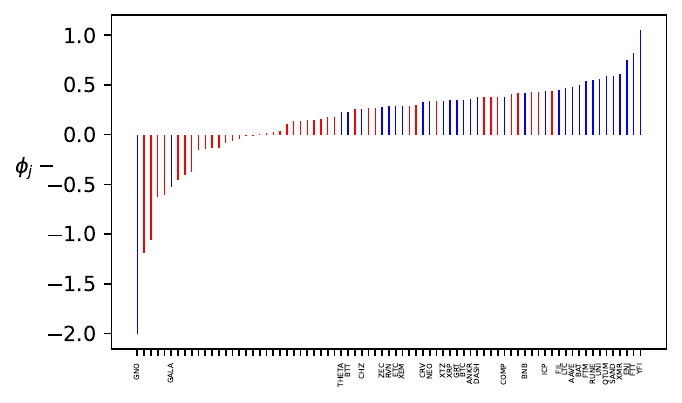}
\end{subfigure}
\begin{subfigure}{.5\textwidth}
  \centering
  \includegraphics[width=\linewidth]{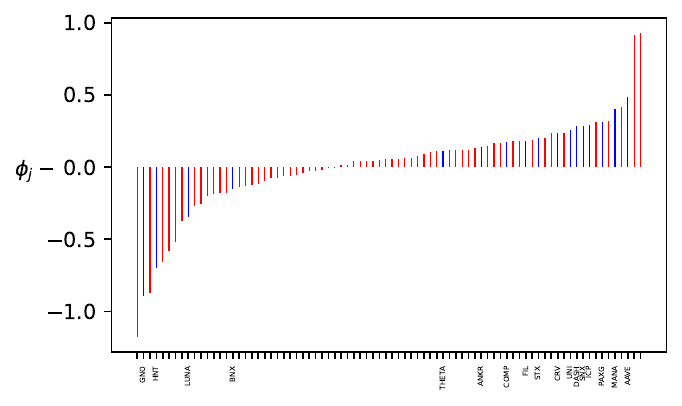}
\end{subfigure}%
\begin{subfigure}{.5\textwidth}
  \centering
  \includegraphics[width=\linewidth]{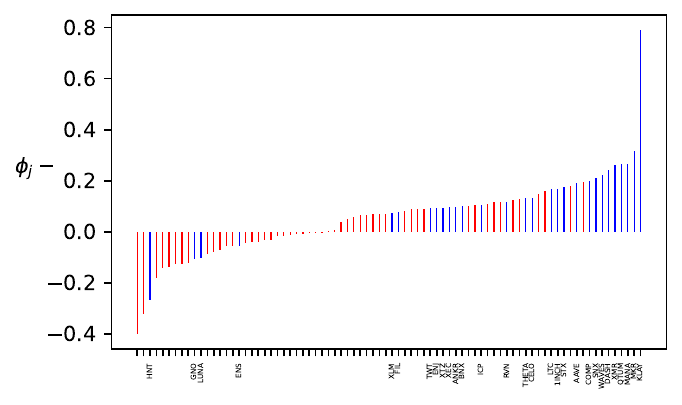}
\end{subfigure}
 \caption{Barplots of $\phi_j^-$ coefficients for the individual HAR-SJV model specification over three different time windows. Each subplot differs for the future horizon of the model specifications: 1 day (top left), 5 days (top right), 22 days (bottom left), and 66 (bottom right). Blue bars identify statistically significant parameters at 5\% confidence level, while red bars are for statistically insignificant ones.}
  \label{Fig:indiv_decSJV_neg}
\end{figure}

\ref{Fig:indiv_decSJV_pos} and \ref{Fig:indiv_decSJV_neg} present the individual estimates for the positive and the negative jump components, whose results are coherent with the panel estimation of the same model specification. Negative jumps exhibit a strong positive effect on future volatility. In contrast, positive jumps are associated with significant estimates mixed among effects of opposite signs in the universe of cryptocurrency analyzed.

\FloatBarrier

\FloatBarrier

\section{Connection to Economic Theories}\label{Sec:connections}

The empirical analysis presented in our study reveals significant insights into the dynamics of cryptocurrency volatility, challenging traditional financial theories with empirical evidence from high-frequency data. Particularly, our findings on the inversion of the traditional leverage effect, where positive returns correspond to increased volatility, suggest a deviation from classical financial theory. This anomaly can be connected to the presence of momentum effects in cryptocurrencies, as reported in \cite{yang2019behavioral}. Established behavioral finance theories, such as those addressing investor overconfidence, herd behavior, and the disposition effect \citep{de1990positive,barberis1998model,hong1999unified,daniel1998investor}, can explain these unique volatility patterns. Cryptocurrencies exhibit a distinct sensitivity to positive returns, possibly driven by the fear of missing out (FOMO), as described in \ref{Subsec:lev} and speculative trading, which are characteristic of overconfident and herd-like investor behavior, following the notion of noise traders in the Glosten-Milgrom model \citep{glosten1985bid}. This divergence underscores the importance of considering psychological and behavioral factors in further analysis of cryptocurrency asset volatility, offering a new perspective on investor behavior in these markets, as done with regards to herding behavior in \cite{da2019herding,gurdgiev2020herding,vidal2019herding,stavroyiannis2019herding}.

The findings from our study on cryptocurrency volatility dynamics align with the principles of market microstructure theory, particularly when viewed through the lens of the AMH theory \citep{lo2004adaptive} as explored by \cite{khuntia2018adaptive}. The cryptocurrency market's continuous trading cycle over the whole day \citep{demiralay2021dynamic} and the prevalence of speculators and uninformed traders \citep{grobys2021speculation} contribute to its dynamic nature, contrasting with more traditional asset classes. In this context, the AMH's emphasis on market adaptation with temporary inefficiency and unusual patterns in price volatility is particularly relevant, as the pace of technological innovation and regulatory developments continually shape cryptocurrency market dynamics. Specifically, the reliance on blockchain technology as the underlying framework for this new asset class introduces new microstructure variables to consider, such as the number of network addresses related to a particular token and the processing power needed to maintain the corresponding blockchain network in place (see \cite{koutmos2018liquidity} for such an analysis on BTC and \cite{fanning2016blockchain} for an evaluation of the technology impact on the financial services industry). The cryptocurrency markets show notable distinctions from traditional markets that can motivate the observed volatility patterns in this work. \cite{silantyev2019order} observe those as coming primarily from shallower average depths of the order book, leading to various disparities in how order books handle order flow. Trade flow imbalances appear effective in explaining price changes in cryptocurrencies.

\section{Discussion}\label{Sec:discussion}

This section evaluates the implications of our findings on the cryptocurrency market's unique volatility landscape, drawing on high-frequency volatility estimators. The econometrics analysis aims to offer valuable insights for market participants and regulators, enriching the understanding of this emerging asset class's volatility patterns and stakeholder roles.

\subsection*{Key Findings}

Our empirical results reveal the cryptocurrency market's volatility dynamics as distinctly marked by speculative behaviors and psychological influences. A primary observation is the inversion of the traditional leverage effect, where positive returns are correlated with future volatility, suggesting a market propelled by speculative enthusiasm and the psychological phenomenon of Fear of Missing Out (FoMo). This inversion signifies a heightened market activity in response to positive gains, attracting investors who anticipate joining a new investment opportunity.

FoMo, particularly pronounced among retail investors, emerges as a crucial determinant of investment behavior, encouraging entry into the market on market downturns as opportunities for purchasing at perceived discounts. This behavior underscores a herding effect in investment decisions, influenced by a collective eagerness to participate in perceived lucrative opportunities without fully grasping the market's complexities. When separately estimating the selected model specification over the three years that compose our dataset, we find that the bearish period in 2022 is the one that is actually driving the inversion of the leverage effect more than the two previous years.

The distinction between investor types, retail versus institutional, further elucidates market dynamics. Retail investors, driven by FoMo and speculative motives, often diverge from traditional market analysis, leading to an inversion in leverage effect dynamics. Conversely, institutional investors adhere to a more conventional understanding of market movements, leveraging price variations for arbitrage and strategic advantage, highlighting their role in sustaining market stability through informed decision-making. In this regard, we document the asymmetric effect of returns when accounting for the intraday jump dynamics. This distinction between investor behaviors underpins the varied volatility drivers in the market. On one side, there are retail investors driven by Fear of Missing Out (FoMO), causing an inversion of the leverage effect at the daily level, and on the other, informed traders exploiting price jumps for arbitrage purposes, causing the usual asymmetric leverage effect at the intraday level.

Moreover, we observe less persistence and volatility memory in cryptocurrency than in traditional equity markets. This finding motivates a deeper investigation into the underlying factors that contribute to these distinctive volatility patterns. The cryptocurrency market, characterized by its innovative technologies and trading mechanisms, diverges significantly from equity markets. The increase in on-chain activity, though still less predominant than off-chain trades, introduces novel financial dynamics not present in traditional finance. Additionally, the cryptocurrency market's structure, which includes both centralized and decentralized platforms, creates unique arbitrage opportunities. These factors lead to a slower assimilation of information into cryptocurrency prices as the market navigates through these dual layers of activity. This complex and evolving landscape underlines the importance of understanding the unique aspects of the cryptocurrency market, particularly its reduced volatility persistence, and memory, shedding light on its distinct behavior and challenges.

\subsection*{Implications and Further Advancements}

The inversion of the traditional leverage effect and the significant influence of FoMo necessitate reconsidering the investment approach to this asset class. This market's distinct behavior, marked by inefficiencies in information transmission and fragmentation across various trading platforms, suggests that the Adaptive Market Hypothesis (AMH) may offer a more accurate framework than the Efficient Market Hypothesis (EMH) for understanding cryptocurrency dynamics. This insight prompts a more cautious and informed approach, acknowledging the psychological underpinnings and the potential for sudden market shifts due to the emergent and still unstable nature of the ecosystem.

The insights derived from our analysis of cryptocurrency market volatility underscore the potential benefits of developing more sophisticated pricing models for cryptocurrency derivatives. While our study does not delve into the exploration of cryptocurrency derivatives directly, we suggest that an in-depth understanding of the unique volatility patterns observed in the cryptocurrency market is crucial for devising accurate risk management strategies. This recognition highlights the importance of derivatives in providing hedging capabilities, necessitating pricing models that effectively account for the cryptocurrency market's distinct volatility characteristics. Consequently, this research points towards the need for innovative derivative pricing mechanisms that can adapt to the peculiar aspects of volatility inherent in cryptocurrencies, thereby aiding in mitigating unexpected volatility patterns. Some preliminary results on cryptocurrency derivatives come from \cite{alexander2023delta}, which analyze the dynamic delta hedging of Bitcoin options, revealing that Bitcoin's implied volatility curves behave very differently from those of equity index options, similarly to the comparison we provide in this article from the historical volatility perspective.

Regulatory efforts should enhance transparency through mandatory disclosure of trading data by exchanges and wallet providers, strengthening financial literacy with targeted educational initiatives on cryptocurrency risks and management. These measures aim to stabilize the market and protect investors against the volatility and speculative risks inherent in cryptocurrency trading. The challenge relies on tailoring oversight mechanisms that recognize and effectively manage the inherent volatility and the complex behaviors of cryptocurrency market participants without harming the increased liquidity and market participation in the cryptocurrency ecosystem. For instance, \cite{chokor2021long} has analyzed the effect of regulation on trader activity, finding that investors reacted less negatively for most illiquid cryptocurrencies and those with higher information asymmetry. However, \cite{feinstein2021impact} notes that such an effect is unclear.

As a recommendation for further advancements in the cryptocurrency ecosystem, we acknowledge that volatility is a crucial component of any derivative pricing model and can benefit the activities of both retail investors and traders. Also, cryptocurrency markets are relatively new and less efficient than traditional financial markets. Accurate volatility estimation can help improve market efficiency by providing more reliable information about cryptocurrencies' risk and return characteristics and allowing portfolio managers to use these assets as an alternative asset class. Furthermore, prompt analysis of the volatility dynamics and a solid comparison with traditional asset classes can guide the regulators in shaping policies for this nascent asset class, which has been the object of an ongoing debate, especially in the last two years. The increasing attention investment in financial technology \citep{kou2021fintech} further augments the importance of having a framework of rules in place.



\section{Conclusions}\label{Sec:conclusion}
\subsection*{Contribution}

We provide an in-depth analysis of price volatility in the cryptocurrency ecosystem at both the aggregate and the individual levels. At the aggregate level, we analyzed the effects of signed volatility, daily leverage effect, and signed jumps on the ecosystem as a whole. In addition, we compared the results obtained on the cryptocurrencies cross-section to a cross-section of stocks, a more mature asset class. The results showed that positive signed volatility, negative daily leverage, and negative signed jumps positively impact the ecosystem's future volatility. The first two results contrast with common stylized facts of financial volatility in more traditional asset classes, signaling a structural difference in the relatively immature cryptocurrency ecosystem. Then, at the individual level, we analyzed the abovementioned effects for each cryptocurrency in the selected sample, retrieving that most of the cryptocurrencies positively impacted the future volatility for positive signed volatility, with a few exceptions. The results also showed that the daily leverage effect tends to be negative for most cryptocurrencies and statistically significant for most cross-section components when the time horizon is larger. We also found that the dataset's most recent year (2022) contributes to highlighting the effect of positively realized semivariances on future volatility. Such an effect contrasts the notion of the asymmetric impact of returns on the realized variance and points to an inverted asymmetry inherent in the cryptocurrency data.

Overall, the analysis of cryptocurrency volatility by using high-frequency data sheds light on specific traits that this nascent market has had since 2020, going from astounding growth to significant drawdowns and disbelief. Identifying the volatility dynamics and comparing it to a more renowned asset class is a crucial aspect to consider for the cryptocurrency ecosystem as a whole. A better assessment of the cryptocurrency market's volatility could stabilize the market by producing a more liquid and reliable derivative market, which nowadays accounts for a few large centralized exchanges such as Binance and various residual exchanges. 

\subsection*{Further Research Endeavors}
An interesting improvement of the volatility analysis could be generated using sophisticated on-chain data, accounting for the blockchain user activity and network size measure where agents are operating. Indeed, the cryptocurrency ecosystem lives on blockchains that produce a lot of data and do not follow the same logic as the traditional financial market. Therefore, it could be that the information is not fully reflected by the price of coins and tokens on centralized exchanges, as there are two alternative and interoperable layers of activities. Moreover, the autoregressive models we employed in our empirical study are all linear in the parameters, neglecting any possible nonlinear relationship in the volatility pattern. This aspect can be further analyzed by extending the model specifications to regime-switching models or employing machine learning techniques that account for nonlinearities as discussed in \cite{sebastiao2021forecasting}. We reserve these points as a study for further analysis since we believe it is possible to deepen the understanding of the cryptocurrency volatility following those lines. 

\section*{Acknowledgements}
The DAREC initiative is funded by donor Teymour H. Farman-Farmaian.

\section*{List of Abbreviations}
\begin{tabular}{p{3cm} p{9cm}}
BTC & Bitcoin \\
DeFi & Decentralized Finance \\
NDTX & NASDAQ Technology Index \\
RV & Realized Variance \\
BV & Bipower Variation \\
IV & Integrated Variance \\
$RV^+$ & Positive Realized Semivariance \\
$RV^-$ & Negative Realized Semivariance \\
API & Application Programming Interface \\
SJV & Jump Variation \\
$SJV^+$ & Positive Signed Jump Variation \\
$SJV^-$ & Negative Signed Jump Variation \\
OLS & Ordinary Least Squares \\
HAR & Heterogeneuous Autoregression \\
WLS & Weighted Least Squares \\
GARCH & Generalized AutoRegressive Conditional Heteroskedasticity \\
GJR-GARCH & Glosten-Jagannathan-Runkle GARCH \\
NFT & Non-Fungible Token \\
FoMO & Fear-of-missing-out \\
ICO & Initial Coin offering
\end{tabular}

\bibliographystyle{dcu}
\bibliography{main_arxiv}

\newpage
\appendix
\section{Additional Estimation Results}
\begin{table}[H]
    \caption*{\small{$RV_{h,t+h} = \mu + \phi_d RV_t +   \phi^-_d RV^-_t + \phi^+_d RV^+_t + \phi^-_w RV^-_t + \phi^+_w RV^+_t +\phi^-_m RV^-_t + \phi^+_m RV^+_t  + \epsilon_{t+h}$}}
\centering
\begin{tabular}{cccccccc} 
\toprule
        &       &  $\mathbf{\phi^-_d}$ &  $\mathbf{\phi^+_d}$  & $\mathbf{\phi^-_w}$ &  $\mathbf{\phi^+_w}$ & $\mathbf{\phi^-_m}$ &  $\mathbf{\phi^+_m}$  \\
\midrule
 \multirow{8}{*}{\rotatebox[origin=c]{90}{Crypto}} & \multirow{2}{*}{\text{t+1}} & -0.108 & 1.224\threeS & -0.001 & 0.152\threeS & -0.245\threeS & 0.472\threeS \\
 && \footnotesize{(-1.311)} & \footnotesize{(9.895)} & \footnotesize{(-0.037)} & \footnotesize{(2.607)} & \footnotesize{(-3.473)} & \footnotesize{(4.218)} \\[1.0ex]
\cmidrule{3-8}
& \multirow{2}{*}{\text{t+5}} & 0.157 & 0.433 & -0.037 & 0.229\threeS & -0.316\threeS & 0.479\threeS \\
 && \footnotesize{(0.425)} & \footnotesize{(1.473)} & \footnotesize{(-0.921)} & \footnotesize{(3.448)} & \footnotesize{(-5.615)} & \footnotesize{(9.886)} \\[1.0ex]                 
\cmidrule{3-8}
 &\multirow{2}{*}{\text{t+22}} & -0.048 & 0.196\threeS & 0.042\twoS & 0.11\threeS & -0.48\threeS & 0.729\threeS \\
 && \footnotesize{(-0.711)} & \footnotesize{(3.56)} & \footnotesize{(2.442)} & \footnotesize{(4.493)} & \footnotesize{(-21.146)} & \footnotesize{(21.917)} \\[1.0ex]      
\cmidrule{3-8}
 &\multirow{2}{*}{\text{t+66}} & -0.054\threeS & 0.108\threeS & 0.02\twoS & 0.07\threeS & -0.37\threeS & 0.528\threeS \\
 && \footnotesize{(-4.333)} & \footnotesize{(6.968)} & \footnotesize{(2.144)} & \footnotesize{(5.252)} & \footnotesize{(-23.174)} & \footnotesize{(23.225)} \\[1.0ex]
                                  
\midrule
  \multirow{8}{*}{\rotatebox[origin=c]{90}{Equity}} &   \multirow{2}{*}{\text{t+1}} & 0.073\twoS & 0.3\threeS & 0.282\threeS & 0.154\threeS & -0.136\oneS & 0.477\threeS \\
 && \footnotesize{(2.277)} & \footnotesize{(5.347)} & \footnotesize{(3.94)} & \footnotesize{(2.321)} & \footnotesize{(-1.971)} & \footnotesize{(5.373)} \\[1.0ex]

\cmidrule{3-8}
   &   \multirow{2}{*}{\text{t+5}}   & 0.054\twoS & 0.12\threeS & 0.184\threeS & 0.258\threeS & -0.167\threeS & 0.504\threeS \\
 & & \footnotesize{(2.416)} & \footnotesize{(4.945)} & \footnotesize{(5.735)} & \footnotesize{(6.884)} & \footnotesize{(-4.252)} & \footnotesize{(10.034)} \\[1.0ex]

\cmidrule{3-8}
  & \multirow{2}{*}{\text{t+22}}   & 0.009 & 0.085\threeS & 0.096\threeS & 0.198\threeS & -0.081\twoS & 0.423\threeS \\
 & & \footnotesize{(0.77)} & \footnotesize{(5.498)} & \footnotesize{(4.546)} & \footnotesize{(7.799)} & \footnotesize{(-2.336)} & \footnotesize{(11.013)} \\[1.0ex]

\cmidrule{3-8}
 &  \multirow{2}{*}{\text{t+66}}   & 0.01 & 0.046\threeS & 0.101\threeS & 0.161\threeS & 0.123\threeS & 0.313\threeS \\
 & & \footnotesize{(1.045)} & \footnotesize{(4.755)} & \footnotesize{(5.706)} & \footnotesize{(9.176)} & \footnotesize{(4.011)} & \footnotesize{(10.867)} \\[1.0ex]
\bottomrule
\multicolumn{3}{l}{\textsuperscript{***}$p<0.01$, 
  \textsuperscript{**}$p<0.05$, 
  \textsuperscript{*}$p<0.1$}
\end{tabular}                        
    \vspace{1em}
    \caption{Estimation results for the full HAR-semiRV model specifications. Results are displayed for the cryptocurrency and the equity panel over the four different horizons. T-stats are in parentheses. The model equation on top of the table reflects all the parameters common to the specification reported in this table. The total number of observations available considering all the entities used for the panel estimate is 47787 for crypto and 17111 for equity}
    \label{tab:fullsemiHAR}
\end{table}

\begin{figure}
\centering
\begin{subfigure}{.5\textwidth}
  \centering
  \includegraphics[width=\linewidth]{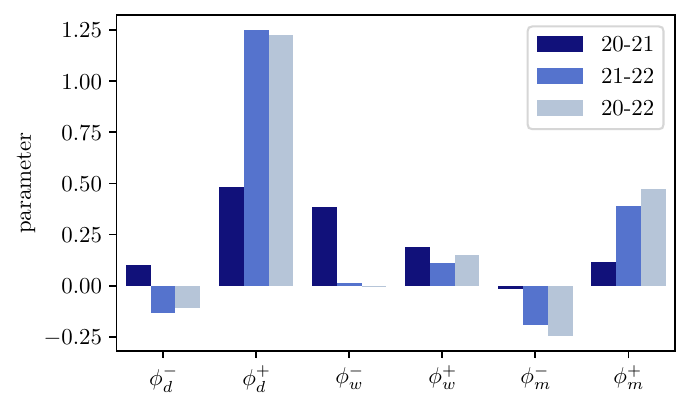}
\end{subfigure}%
\begin{subfigure}{.5\textwidth}
  \centering
  \includegraphics[width=\linewidth]{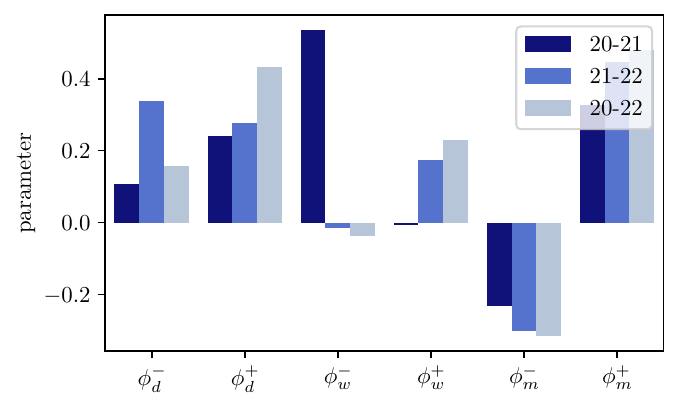}
\end{subfigure}
\begin{subfigure}{.5\textwidth}
  \centering
  \includegraphics[width=\linewidth]{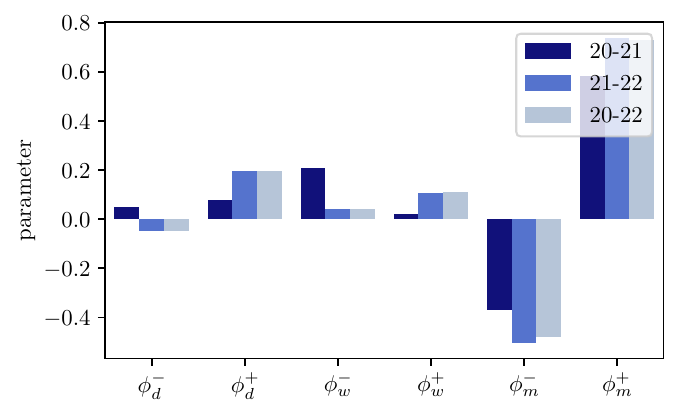}
\end{subfigure}%
\begin{subfigure}{.5\textwidth}
  \centering
  \includegraphics[width=\linewidth]{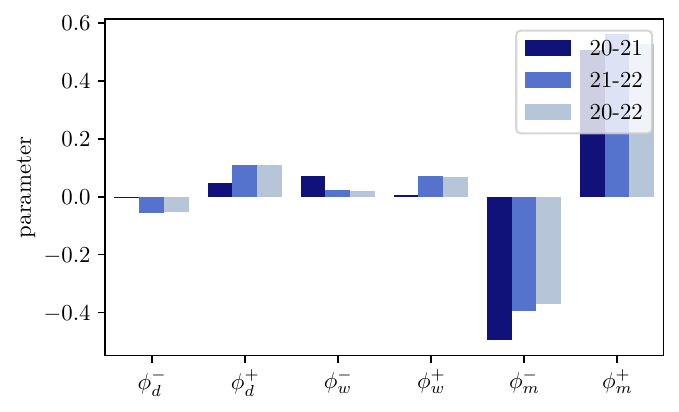}
\end{subfigure}
 \caption{Barplots of estimated coefficient for the full panel HAR-semiRV model specification over three different time windows. Each subplot differs for the future horizon of the model specifications: 1 day (top left), 5 days (top right), 22 days (bottom left), and 66 (bottom right).}
  \label{Fig:DecRV_trends}
\end{figure}%
\FloatBarrier

\begin{table}[H]
    \caption*{\small{$RV_{h,t+h} = \mu +  \phi_c BV_t + \phi_w RV_{w,t} + \phi_m RV_{m,t} + \epsilon_{t+h}$}}
\centering
\begin{tabular}{ccccc} 
\toprule
        &       &  $\mathbf{\phi_c}$  & $\mathbf{\phi_w}$ & $\mathbf{\phi_m}$  \\
\midrule
 \multirow{8}{*}{\rotatebox[origin=c]{90}{Crypto}} & \multirow{2}{*}{\text{t+1}} & 0.634\threeS & 0.005 & 0.053\threeS \\
 & & \footnotesize{(7.65)} & \footnotesize{(1.35)} & \footnotesize{(4.26)} \\[1.0ex]
\cmidrule{3-5}
& \multirow{2}{*}{\text{t+5}} & 0.318\threeS & 0.051\twoS & 0.05\threeS \\
 & & \footnotesize{(5.445)} & \footnotesize{(2.525)} & \footnotesize{(3.661)} \\[1.0ex]           
\cmidrule{3-5}
 &\multirow{2}{*}{\text{t+22}} & 0.09\threeS & 0.028\threeS & 0.048\threeS \\
 & & \footnotesize{(4.021)} & \footnotesize{(2.58)} & \footnotesize{(4.084)} \\[1.0ex]    
\cmidrule{3-5}
 &\multirow{2}{*}{\text{t+66}} & 0.03\threeS & 0.006\oneS & -0.012\threeS \\
 & & \footnotesize{(2.651)} & \footnotesize{(1.188)} & \footnotesize{(-2.721)} \\[1.0ex]
                                  
\midrule
  \multirow{8}{*}{\rotatebox[origin=c]{90}{Equity}} &   \multirow{2}{*}{\text{t+1}} & 1.088\threeS & 0.126\threeS & 0.129\threeS \\
 & & \footnotesize{(9.098)} & \footnotesize{(5.202)} & \footnotesize{(2.588)} \\[1.0ex]

\cmidrule{3-5}
   &   \multirow{2}{*}{\text{t+5}}    & 0.582\threeS & 0.148\threeS & 0.14\threeS \\
 & & \footnotesize{(11.173)} & \footnotesize{(10.17)} & \footnotesize{(5.491)} \\[1.0ex]

\cmidrule{3-5}
  & \multirow{2}{*}{\text{t+22}}   & 0.335\threeS & 0.087\threeS & 0.154\threeS \\
 & & \footnotesize{(9.397)} & \footnotesize{(8.592)} & \footnotesize{(11.512)} \\[1.0ex]

\cmidrule{3-5}
 &  \multirow{2}{*}{\text{t+66}}    & 0.209\threeS & 0.067\threeS & 0.205\threeS \\
 & & \footnotesize{(8.755)} & \footnotesize{(9.528)} & \footnotesize{(22.162)} \\[1.0ex]
\bottomrule
\multicolumn{3}{l}{\textsuperscript{***}$p<0.01$, 
  \textsuperscript{**}$p<0.05$, 
  \textsuperscript{*}$p<0.1$}
\end{tabular}                        
    \vspace{1em}
    \caption{Estimation results for HAR-SJV model specifications with no jump components at the first lag. Results are displayed for the cryptocurrency and the equity panel over the four different horizons. T-stats are in parentheses. The model equation on top of the table reflects all the parameters common to the specification reported in this table. The total number of observations available considering all the entities used for the panel estimate is 47787 for crypto and 17111 for equity}
    \label{tab:HAR_NOjumps}
\end{table}
\FloatBarrier

\begin{figure}
\centering
\begin{subfigure}{.5\textwidth}
  \centering
  \includegraphics[width=\linewidth]{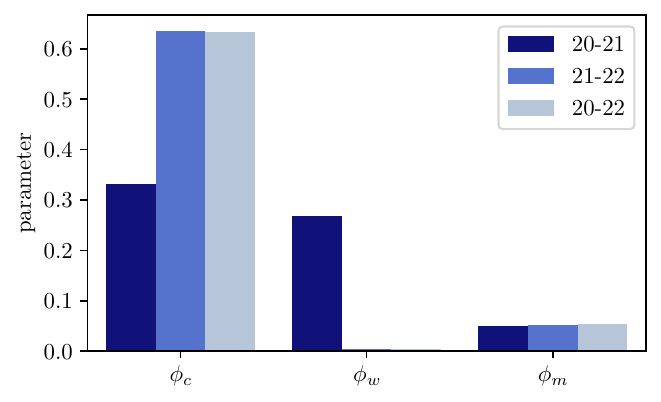}
\end{subfigure}%
\begin{subfigure}{.5\textwidth}
  \centering
  \includegraphics[width=\linewidth]{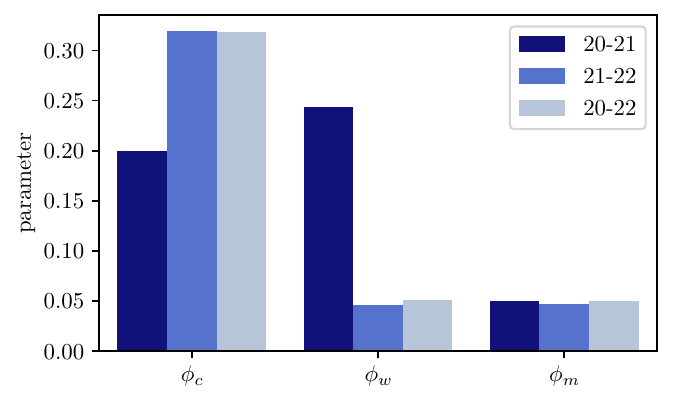}
\end{subfigure}
\begin{subfigure}{.5\textwidth}
  \centering
  \includegraphics[width=\linewidth]{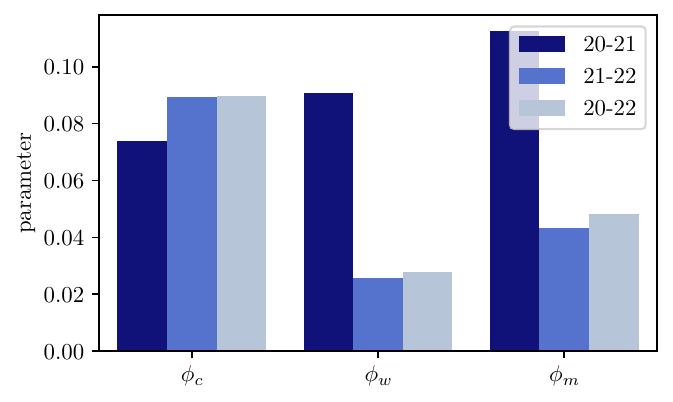}
\end{subfigure}%
\begin{subfigure}{.5\textwidth}
  \centering
  \includegraphics[width=\linewidth]{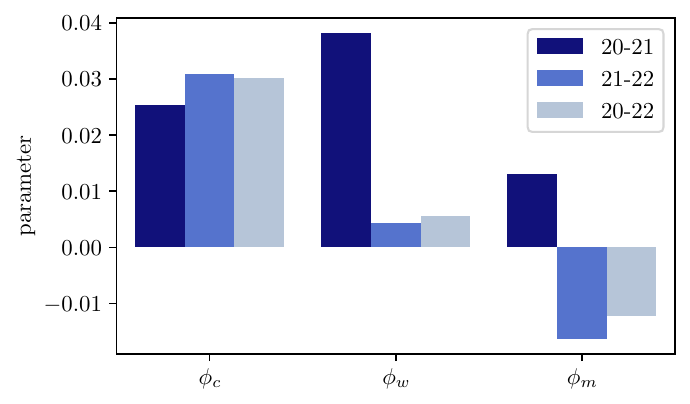}
\end{subfigure}
 \caption{Barplots of estimated coefficient for the panel for HAR-SJV model specifications with no jump components over three different time windows. The number in the legend reflects the year of analysis and is inclusive. Each subplot differs for the future horizon of the model specifications: 1 day (top left), 5 days (top right), 22 days (bottom left), and 66 (bottom right).}
  \label{Fig:BVnojump_trends}
\end{figure}%
\FloatBarrier

\end{document}